\documentclass[a4paper,11pt]{article}
\pdfoutput=1 
\usepackage[english]{babel}
\usepackage{microtype,xspace}
\usepackage[utf8]{inputenc}	

\usepackage{jheppub} 
\usepackage{bm,amssymb,amsmath,slashed,graphicx,multirow,soul,mathtools,xspace,array}
\usepackage{siunitx}
\usepackage{bm} 
\usepackage{float}
\usepackage{feynmf}
\allowdisplaybreaks
\usepackage{ bbold }
\usepackage{subfigure}
\usepackage[normalem]{ulem} 

\newcommand{\todo}[1]{{\color{red} \ifmmode\else[todo]\fi #1}}
\usepackage[usenames,dvipsnames]{xcolor}
     \definecolor{hgreen}{rgb}{0,.3,0}
      \definecolor{darkgreen}{rgb}{0.3,.8,0.2}
     \definecolor{hred}{rgb}{.3,0,0}
     \definecolor{hblue}{rgb}{0,0,.3}
     \definecolor{LightGray}{gray}{0.95}

\newcommand{\TeV}{\text{TeV}}
\newcommand{\GeV}{\text{GeV}}
\newcommand{\Br}{\text{Br}}

\newcommand{\beq}{\begin{equation} }
\newcommand{\eeq}{\end{equation}}
\newcommand{\be}{\begin{equation} }
\newcommand{\ee}{\end{equation}}
\newcommand{\bi}{\begin{itemize} }
\newcommand{\ei}{\end{itemize} }

\renewcommand{\L}{\mathcal{L}}

\newcommand{\RK}{$ R_{K^{(\ast)}} $\xspace}

\definecolor{Red}{rgb}{1.,0.,0.}
\definecolor{Grn}{rgb}{0.,0.75,0.}
\definecolor{Blu}{rgb}{0.,0.,1.}

\definecolor{red}{rgb}{0.6,.0706,.1373}
\definecolor{blue}{rgb}{0,0.396,0.741}

\let\Re\relax
\DeclareMathOperator{\Re}{Re}
\let\Im\relax
\DeclareMathOperator{\Im}{Im}

\usepackage[T1]{fontenc} 

\usepackage{amsmath,amssymb,epsfig,color,slashed}

\allowdisplaybreaks

\title{\boldmath New Physics in $b \to s \mu \mu$: FCC-hh or a Muon Collider?}

\author[1,3]{Aleksandr Azatov,}
\author[1,3]{Francesco Garosi,}
\author[2]{Admir Greljo,}
\author[3]{David Marzocca,}
\author[2]{Jakub Salko,}
\author[3]{Sokratis Trifinopoulos}

\affiliation[1]{SISSA International School for Advanced Studies, Via Bonomea 265, 34136, Trieste, Italy}
\affiliation[2]{Albert Einstein Center for Fundamental Physics, Institute for Theoretical Physics, University of Bern, CH-3012 Bern, Switzerland}
\affiliation[3]{INFN, Sezione di Trieste, SISSA, Via Bonomea 265, 34136, Trieste, Italy}

\abstract{Rare flavour-changing neutral-current transitions $b \to s \mu^+ \mu^-$ probe higher energy scales than what is directly accessible at the LHC. Therefore, the presence of new physics in such transitions, as suggested by the present-day LHCb anomalies, would have a major impact on the motivation and planning of future high-energy colliders. 
The two most prominent options currently debated are a proton-proton collider at 100 TeV (FCC-hh) and a multi-TeV
muon collider (MuC). In this work, we compare the discovery prospects at these colliders on benchmark new physics models indirectly detectable in $b \to s \mu^+ \mu^-$ decays but beyond the reach of the high-$p_T$ searches at the HL-LHC. We consider a comprehensive set of scenarios: semileptonic contact interactions, $Z^\prime$ from a gauged $U(1)_{B_3 - L_\mu}$ and $U(1)_{L_\mu - L_\tau}$, the scalar leptoquark $S_3$, and the vector leptoquark $U_1$. We find that a $3~\TeV$ MuC has a sensitivity reach comparable to the one of the FCC-hh. However, for a heavy enough mediator, the new physics effects at a 3 TeV MuC are only observed indirectly via deviations in the highest energy bin, while the FCC-hh has a greater potential for the discovery of a resonance. Finally, to completely cover the parameter space suggested by the $bs\mu\mu$ anomalies, among the proposed future colliders, only a MuC of 10 TeV (or higher) can meet the challenge.
}

\begin{document}

\maketitle

\flushbottom

\newpage

\section{Introduction}
\label{sec:intro}

Ten years have passed since the discovery of the Higgs boson at the Large Hadron Collider (LHC) by the ATLAS and CMS collaborations~\cite{ATLAS:2012yve,CMS:2012qbp}. The precision Higgs measurements which followed all agree quite well with the Standard Model (SM) predictions. In the meantime, the two collaborations have searched very hard for a plethora of hypothetical new physics (NP) particles at the energy frontier. This tremendous effort was so far unsuccessful, confirming, once again, the SM in spite of naturalness arguments that suggested the presence of new physics at (or below) the TeV scale. The most compelling discrepancy between LHC experiments and theory nowadays is actually observed in a dedicated flavour physics experiment. The LHCb collaboration unexpectedly found evidence for lepton flavour universality (LFU) violation~\cite{LHCb:2017avl,LHCb:2021trn} which, if correct, would not only represent a long-sought breakdown of the SM close to the TeV scale but would also point towards some exotic beyond the SM (BSM) scenario.

The anomalies in several $B$-meson decays can be coherently explained by a short-distance new physics contribution in the underlying quark-level transition $b \to s \mu^+ \mu^-$ (we refer to this as $bs\mu\mu$ anomalies in what follows). Provided this effect is not due to a yet unknown experimental systematics, we are led to conclude that there exists a new \textit{super-weak} Fermi force of $\mathcal{O}(10^{-5}) \,G_F$. In analogy with the prediction of the weak scale from $G_F$, this gives us information about the scale where new mediator states are integrated out. The violation of perturbative unitarity in $2 \to 2$ scattering implies a new state with mass below (roughly) $100$\,TeV~\cite{DiLuzio:2017chi}. In other words, if the observed anomalies are indeed due to new physics and the basic postulates of quantum field theory hold, we have empirical evidence for a new mass threshold in the vicinity of our present and planned colliders.

While a final word on the flavour anomalies (due to either new physics or experimental systematics) might take several years or even a decade,  discussions about building new colliders are already taking place \cite{Adolphsen:2022bay}, with the 100 TeV future circular hadron collider (FCC-hh) \cite{FCC:2018byv,FCC:2018vvp,Bernardi:2022hny} and multi-TeV muon collider (MuC) \cite{Stratakis:2022zsk,Jindariani:2022gxj,Aime:2022flm,DeBlas:2022wxr} being the most promising options for the energy frontier.\footnote{Instead, the next generation of EW and Higgs factories will explore the precision frontier, with FCC-ee~\cite{FCC:2018evy}, ILC~\cite{Brau:2007zza}, and CLIC~\cite{CLIC:2018fvx} as the most promising proposals in this regard.} If the present-day flavour anomalies are due to new physics, this would provide a clear target for direct discoveries at future colliders. Conversely, it is still important to investigate the complementary high-$p_T$ searches at future colliders and compare them with the indirect reach from flavour physics. In this work, we will thoroughly explore a few motivated new physics benchmark scenarios relevant for (semi)leptonic $B$-meson decays: $i)$ flavourful four-fermion interactions, $ii)$ gauged $U(1)$ extensions and $iii)$ leptoquark models. We will assess and compare the prospects at colliders listed in Table~\ref{tab:colliders}. The current LHC bounds and the HL-LHC projections define the target for the next generation of colliders. Our main goal is to compare FCC-hh versus the MuC in the leftover parameter space.

The physics of the MuC relevant for this work might be somewhat less familiar to the reader and is discussed already in Section~\ref{sec:MuC}. Section~\ref{sec:hadron} reviews the complementary signatures at hadron colliders.

\begin{table}[t]
\begin{center}
\begin{tabular}{ | c | c | c | c | }
\hline
\textbf{Collider} & \textbf{C.o.m. Energy} & \textbf{Luminosity} & \textbf{Label} \\
\hline
LHC Run-2 & 13 TeV & 140 fb$^{-1}$ & LHC \\
HL-LHC & 14 TeV & 6 ab$^{-1}$ & {\color{magenta} HL-LHC} \\
FCC-hh & 100 TeV & 30 ab$^{-1}$ & {\color{violet} FCC-hh} \\
Muon Collider & 3 TeV & 1 ab$^{-1}$ & {\color{Red} MuC3} \\
Muon Collider & 10 TeV & 10 ab$^{-1}$ & {\color{darkgreen} MuC10} \\
Muon Collider & 14 TeV & 20 ab$^{-1}$ & {\color{Blu} MuC14} \\
\hline
\end{tabular}
\caption{The energy and the luminosity of benchmark colliders. The detector specifications for FCC-hh and MuC are discussed in Appendix~\ref{app:Detector}. The last column shows the short-hand label and color code for each collider, used for all the sensitivity plots in the paper.\label{tab:colliders}}
\end{center} 
\end{table}

In Section~\ref{sec:contact} we consider the most pessimistic scenario, where the NP states are too massive to be produced on-shell, and then look for the correlated effect of the new semileptonic contact interactions in the high-energy tails. As an example, we consider four-fermion operators composed of two quarks and two second-generation leptons, all $SU(2)_L$ doublets. We only assume couplings to muons for a more direct comparison between MuC and FCC-hh and due to the additional motivation of the $bs\mu\mu$ anomalies.
Firstly, we consider a minimal flavour violation (MFV)~\cite{DAmbrosio:2002vsn} scenario, where the leading EFT coefficients are proportional to the identity in quark flavour space.\footnote{In Ref.~\cite{Greljo:2017vvb} it was shown that this scenario is disfavoured as solution to $bs\mu\mu$ anomalies due to tension with LHC constraints from $p p \to \mu^+ \mu^-$.}
Then, as a scenario more related to $bs\mu\mu$ anomalies, we consider a minimally broken $U(2)^3$ flavour symmetry~\cite{Barbieri:2011ci,Faroughy:2020ina,Greljo:2022cah}. We compare the high-$p_T$ bounds with the tentative values suggested by the flavour anomalies. 

After deriving limits on the contact interactions, we study explicit mediator models, focusing only on tree-level mediators.
There is a finite number of tree-level mediators which can produce a semileptonic four-fermion interaction at low energies.  These are bosons, either color singlets or triplets.
We restrict our discussion to the $Z'$ (color-singlet vectors) and leptoquarks (color-triplet scalars or vectors).

In Section~\ref{sec:Zprime} we study two representative examples of the $Z'$ models, where the mediator is a massive gauge boson of a spontaneously broken gauge symmetry $U(1)_{B_3 - L_\mu}$ (in Section~\ref{sec:ZprimeB3Lmu}) and $U(1)_{L_\mu - L_\tau}$ (in Section~\ref{sec:ZprimeLmuLtau}), respectively. Both are free of chiral anomalies with the fermionic content minimally extended to include right-handed neutrinos. The first example represents a large class of models in which the $Z'$ interaction with the third generation of quarks is of the same size as the one with muons, while the interactions with the light quark families are suppressed. This model is motivated by the flavour structure observed in the SM quark sector which has an approximate $U(2)^3$ flavour symmetry. The second example represents models in which quark interactions are altogether suppressed, compared with those to muons. Both classes of models are less constrained from current LHC data, compared with the quark-universal $Z'$ models with $B/L_\mu \sim \mathcal{O}(1)$ that are produced from valence quarks.
In this work, we focus on the $Z'$ mass range above the electroweak scale. For both models, we first assume only the renormalisable couplings present in the unbroken $U(1)_X$ phase (we assume the breaking by the condensate of a SM-singlet scalar).
Then, we switch on also the minimal set of other couplings required to fit the $bs\mu\mu$ anomalies and impose that they are addressed by the model.

In Section~\ref{sec:LQ}, we consider separately two single-leptoquark simplified models: a scalar weak triplet $S_3$ (in Section~\ref{sec:LQscalar}) and a vector weak singlet $U_1$ (in Section~\ref{sec:LQvector}), where we use the nomenclature of Ref.~\cite{Dorsner:2016wpm}.
In each case, first we assume that the flavour structure of the coupling matrix respects the $U(2)^3$ quark flavour symmetry, allowing for the interactions with the third generation of quarks only.
We also assume that the leptoquark carries a muon number, thus coupling only to muons. For $S_3$, this condition naturally emerges in the context of lepton-flavour non-universal $U(1)_X$ gauge extensions under which the leptoquark is charged such that it has conserved baryon and muon numbers~\cite{Greljo:2021xmg,Greljo:2021npi,Davighi:2020qqa,Hambye:2017qix,Davighi:2022qgb,Heeck:2022znj}. Then, in the study specific to the $bs\mu\mu$ anomalies, we assume the minimal set of couplings needed to fit the latest data consistent with the $SU(2)_L$ gauge invariance. Besides the $b \mu $-LQ coupling, already present in the previous scenario, this set necessarily includes also the $s \mu$-LQ coupling. The ultraviolet complete models with the vector leptoquark however will necessarily have additional states close in mass to the leptoquark increasing the chance for discovery~\cite{DiLuzio:2017vat,Greljo:2018tuh,Bordone:2017bld,Bordone:2018nbg,Cornella:2019hct,Fornal:2018dqn,Blanke:2018sro,Fuentes-Martin:2019ign,Guadagnoli:2020tlx,Heeck:2018ntp,Fuentes-Martin:2020bnh,Fuentes-Martin:2019bue,Fuentes-Martin:2020luw,Fuentes-Martin:2020hvc,Baker:2021llj,Houtz:2022fnk}. We neglect those states in this study. Finally, since leptoquarks are colored and efficiently produced in hadron colliders, the interesting mass range is set by the direct searches at the LHC, $m_{{\rm LQ}}\gtrsim 1$~TeV.

The reason for choosing diverse benchmark models is to broadly cover the BSM space. There have already been several prospect studies for some of these models at the FCC-hh~\cite{Allanach:2017bta,Allanach:2018odd,Allanach:2019zfr,Garland:2021ghw,Helsens:2019bfw,Bandyopadhyay:2020wfv,Bandyopadhyay:2021pld}, MuC~\cite{Asadi:2021gah,Huang:2021nkl,Huang:2021biu,Qian:2021ihf,Bandyopadhyay:2021pld,Capdevilla:2021rwo,MuonCollider:2022xlm,Altmannshofer:2022xri}, CLIC~\cite{CLIC:2018fvx} and HL(HE)-LHC~\cite{Allanach:2017bta,Allanach:2018odd,Cerri:2018ypt,CidVidal:2018eel}. Throughout the paper we comment on the similarities and differences between our study and some of the previous ones. An important novelty is the use of the muon parton distribution functions (PDFs), of which some results are presented in Appendix~\ref{app:MuonPDFs} and further details will be published in a dedicated work \cite{muPDF}.
In Appendices~\ref{app:partonic_xsec} and \ref{app:statistics} we report some analytic results for cross sections at MuC and details of our statistical analysis.
In Appendix~\ref{app:Detector} we discuss our assumptions about the future detector performance and finally in Appendix~\ref{app:FCChh} we present details of our numerical studies for hadron colliders. The conclusions are given in Section~\ref{sec:conc}.

\begin{figure}[t]
\centering
\begin{tabular}{cccc}
  \multicolumn{2}{c}{$\mu^+ \mu^- \to j j$} &
  \multicolumn{2}{c}{$\mu^+ \mu^- \to \mu^+ \mu^-$} \\
  \includegraphics[height=3cm]{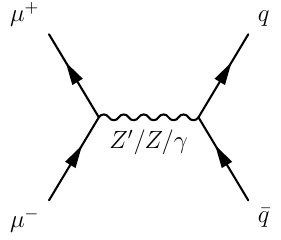} &   
  \includegraphics[height=3cm]{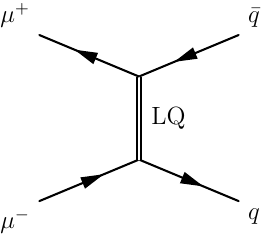} &
  \includegraphics[height=3cm]{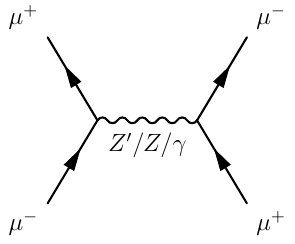} &
\includegraphics[height=3cm]{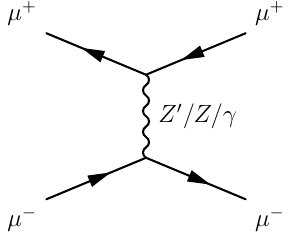} \\[0.4cm]
  \multicolumn{2}{c}{$\mu q \to \mu j$} &
  \multicolumn{2}{c}{$\mu^+ \mu^- \to {\rm LQ} \overline{{\rm LQ}}$} \\
  \includegraphics[width=3cm]{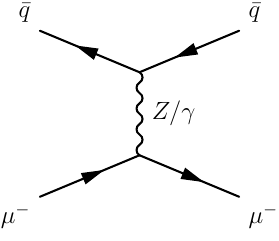} &
  \includegraphics[width=3cm]{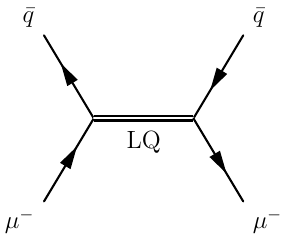} &
  \includegraphics[width=3cm]{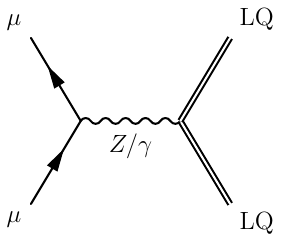} &
  \includegraphics[width=3cm]{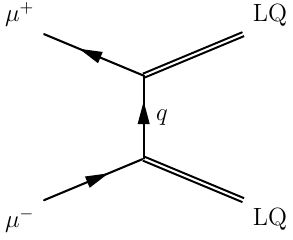}
\end{tabular}
\caption{\label{fig:FDiagrams} Feynman diagrams for the partonic processes relevant for our MuC phenomenology. For the scalar leptoquark $S_3$ one should exchange $q \leftrightarrow \bar{q}$.}
\end{figure}

\section{Signatures at a muon collider}
\label{sec:MuC}

Muon colliders combine the advantages of both proton-proton and electron-positron colliders: high energy reach, where all the collider energy is accessible in $\mu^+ \mu^-$ collisions, with high precision measurements, thanks to the low QCD background and clean initial state \cite{Buttazzo:2020uzc,AlAli:2021let,Aime:2022flm,DeBlas:2022wxr}.

As for proton collisions, also for high-energy MuC it is important to take into account collinear radiation emitted by splitting of the initial state. In both colliders these processes can be described in terms of parton distribution functions. In case of muons (the same holds also for electrons or positrons) for energies below the electroweak scale the main effect is due to QED interaction, with QCD providing a secondary, but still important, contribution: muons emit photons, that can split into $q\bar{q}$ or $\ell^+ \ell^-$ pairs, that can emit gluons or photons, etc. Above the EW scale, instead, QED is substituted by EW interactions, which thus become the dominant effect. This is in contrast to a proton collider, where QCD is the dominant interaction both below and above the EW scale.
In the evaluation of the physics potential of MuC we therefore take into account the complete EW PDF of muons, see App.~\ref{app:MuonPDFs} and Refs.~\cite{Han:2020uid,Han:2021kes,muPDF} for recent results on the subject. We denote the muon (anti-muon) beam as $\bm{\mu}$ ($\bm{\bar \mu}$), while the individual partons are $\mu^\pm, \ell^\pm, \nu_i, q_i, \bar{q}_i, \gamma, W, Z,$ etc.

For the $Z^\prime$ and leptoquark benchmark models, the relevant MuC processes are: di-jet and di-tau production from muon annihilation ($\mu^+ \mu^- \to j j, \tau^+ \tau^-$), Bhabha scattering of muons ($\mu^+ \mu^- \to \mu^+ \mu^-$), muon-quark scattering ($\mu^{\pm} q \to \mu^{\pm} q$, that includes single production of leptoquark), and pair production of leptoquarks ($\mu^+ \mu^- \to {\rm LQ} \overline{{\rm LQ}}$).

Except for $\mu q \to \mu q$, all the other processes we study are initiated by $\mu^- \mu^+$, i.e. the valence partons inside the muonic and anti-muonic beam, respectively.
The $\mu^- \mu^+$ luminosity $\mathcal{L}_{\mu\mu}(m_{\mu\mu})$ grows when $m_{\mu\mu} \to 0$ (see Fig.~\ref{fig:pdfsmub} in App.~\ref{app:MuonPDFs}) due to the contribution arising from the splitting of photons and EW gauge bosons, as well as when going closer to the collider energy  $m_{\mu\mu} \to \sqrt{s_0}$, with a minimum in the intermediate energies.
This behavior is completely different than $q -\bar{q}$ luminosities (cf. Eq.~\eqref{eqA:lumin}) in proton-proton colliders, where the luminosity monotonously decreases going to higher energies  and becomes negligible well before the kinematical limit of the collider.
This difference is important to understand our numerical results.
In a MuC, if the NP has a mass below the collider energy one can look for its effect both in the shape of the cross section (a resonance peak or a $t$($u$)-channel exchange) for $m_{\mu\mu} < \sqrt{s_0}$ as well as in the very precise measurement of the cross section at the highest invariant mass bin, $m_{\mu\mu} \approx \sqrt{s_0}$. The latter method works much better at MuC compared to similar methods at hadron colliders, see e.g.~\cite{Becciolini:2014lya,Alves:2014cda}, thanks to the large parton luminosity, lower theory uncertainties, and cleaner collider environment. For NP states heavier than $\sqrt{s_0}$, instead, the sensitivity arises only from the latter strategy.

In the following we provide more details for each of the MuC processes we studied. The differential cross sections are derived after computing analytically the partonic cross sections of the $2\to 2$ processes (see App.~\ref{app:partonic_xsec}) and convoluting them with the parton luminosities of the initial state (see App.~\ref{app:MuonPDFs}).

\subsection{Di-jet: {\boldmath$\mu \bar{\mu}$}$\to j j$}
\label{sec:di-jet}

\begin{figure}[t]
\centering
\includegraphics[height=5.4cm]{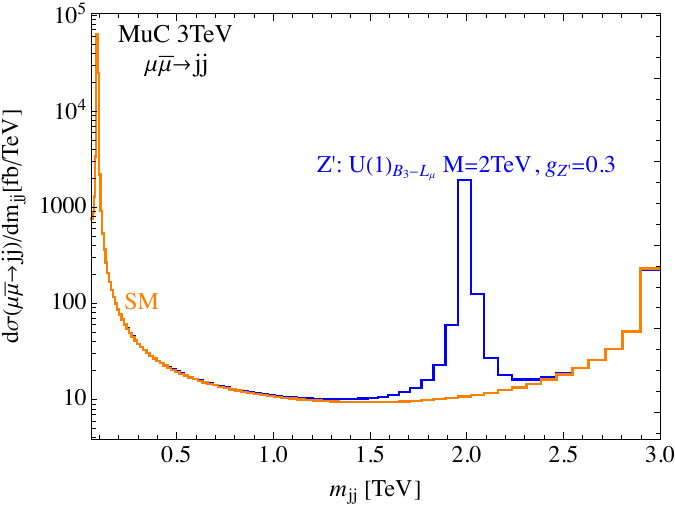} ~
\includegraphics[height=5.3cm]{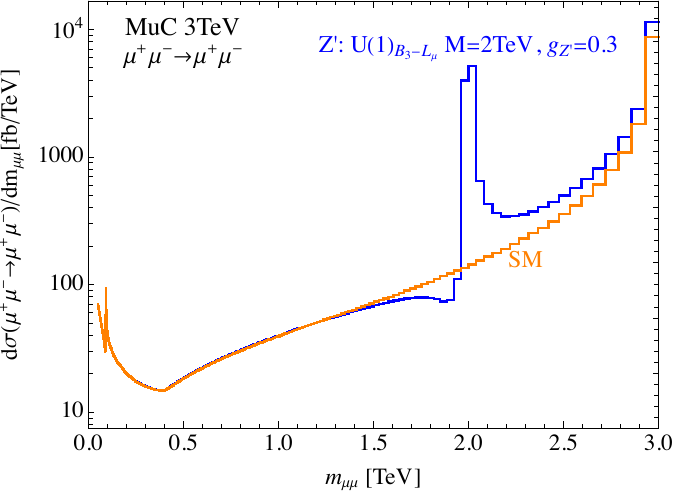} \\
\caption{\label{fig:ZprimeXsecMuC} Differential cross sections for inclusive IDY {\boldmath$\mu \bar{\mu}$}$\to j j$ (left) and $\mu^+ \mu^- \to \mu^+ \mu^-$ (right) for a 3 TeV MuC, taking into account muon PDFs. The SM cross section is shown in orange, while the prediction for a $Z^\prime$ resonance in the $U(1)_{B_3 - L_\mu}$ model with $m_{Z'} = 2$~TeV and coupling $g_{Z^\prime} = 0.3$ is shown in blue.}
\end{figure}

This process is dominated by the \emph{Inverted Drell-Yan} (IDY) channel $\mu^+ \mu^- \to j j$ shown in Fig.~\ref{fig:FDiagrams}.
Due to the non-negligible muon neutrino PDF in the muon beam (see Fig.~\ref{fig:pdfsmub}), a sub-leading but relevant contribution is induced by the charged-current channel $\nu_\mu \mu^+ \to j j$ and its conjugate, while the purely neutrino-induced channel $\nu_\mu \bar{\nu}_\mu \to j j$ is suppressed by the $\nu_\mu \bar{\nu}_\mu$ luminosity, that is always a factor of at least $\sim 10$ smaller than the $\mu^+ \mu^-$ one.
In light of this, in our analysis of the di-jet channel we include both $\mu \mu \to j j$ and $\nu_\mu \mu \to j j$, but neglect the purely neutrino-induced process.
On the other hand, the QCD contribution $q(g) \bar q(g) \to j j$ is always negligible due to the quark and gluon PDF suppression, see App.~\ref{app:MuonPDFs}.\footnote{The $gg$ luminosity is the largest among colored parton combinations at smaller energies, while the $u\bar{u}$ luminosity takes over at higher energies. Nonetheless, both are negligible when compared with the $\mu \bar{\mu}$ luminosity. For $\sqrt{s_0} = 3 \,\TeV$\,collider, the $gg$ luminosity is only $\sim 10^{-2}$ of the $\mu \bar{\mu}$ one at $m_{jj} = 100$\,GeV and decreases further for higher energies,  while $u\bar{u} / \mu \bar{\mu}$ luminosity ratio is $\sim 10^{-4} (10^{-7}$) at 0.5 (2) TeV.}

\begin{figure}[t]
\centering
\includegraphics[height=5.4cm]{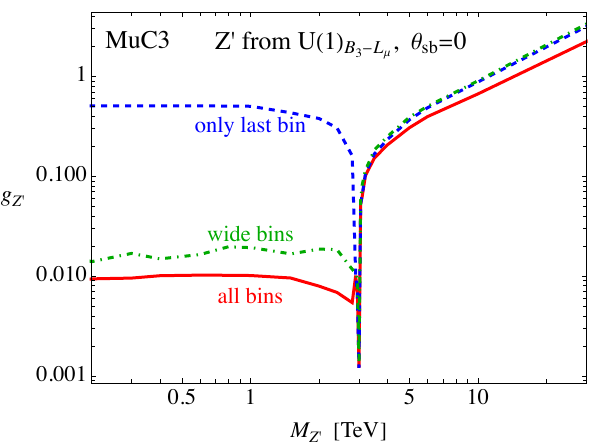} \\
\caption{\label{fig:ZprimeB3Lmu_mumujj} Comparison of 5$\sigma$ discovery prospects for a $Z^\prime$ in a gauged $U(1)_{B_3-L_\mu}$ model (see Section~\ref{sec:ZprimeB3Lmu}) at MuC3 from {\boldmath$\mu \bar{\mu}$}$\to j j$ using the full di-jet resolution (solid red), a few large bins (dot-dashed green), and only the last invariant mass bin (dashed blue).}
\end{figure}

Muonic IDY is very sensitive to new physics coupled to muons and quarks, as in our benchmark models.
After integrating over angular distributions, we construct the di-jet invariant mass bins following the hadronic calorimeter resolution described in App.~\ref{app:resolution} (see Eq.~\eqref{eq:hCalRes}).
In Fig.~\ref{fig:ZprimeXsecMuC} (left) we show the SM IDY cross section for a 3 TeV MuC (orange) as well as the contribution from a $Z^\prime$ resonance (blue) (for details on the specific model see Section~\ref{sec:Zprime}).
We observe that, due to the $\mu^+ \mu^-$ PDF luminosity (see App.~\ref{app:MuonPDFs}) and the shape of the SM partonic cross section, the convoluted SM cross section decreases above the $Z$ boson invariant mass, to then increase again when the $m_{jj}$ invariant mass approaches the collider energy $\sqrt{s_0}$. This behaviour of the SM cross section has important implications for the new physics searches. In case of a four-fermion contact interaction, the strongest sensitivity is obtained from the last few bins, where both the cross section and the energy are largest. For $s$-channel resonance searches, instead, the peak typically provides the dominant sensitivity. In particular, the bulk of the sensitivity is given not so much from the shape of the (possibly narrow) peak as from its integrated contribution to the cross section. Finally, a non-negligible contribution comes from the precision measurement of the cross section in the highest-energy bin. In Fig.~\ref{fig:ZprimeB3Lmu_mumujj} we compare the constraints on a $Z^\prime$ resonance from a gauged $U(1)_{B_3-L_\mu}$ model (see Section~\ref{sec:ZprimeB3Lmu} for details) at MuC3 from {\boldmath$\mu \bar{\mu}$}$\to j j$. The solid red line is obtained employing the full calorimeter resolution, the dot-dashed green prospect instead uses a few bins much larger than the resonance width ([0.15-0.3-0.7-2-3] TeV), and finally the dashed blue prospect is obtained by considering only the very last invariant mass bin ([2.9-3] TeV). This also shows that the specific choice we employ for the calorimeter resolution does not affect much our results.

\begin{figure}[t]
\centering
\includegraphics[height=5.0cm]{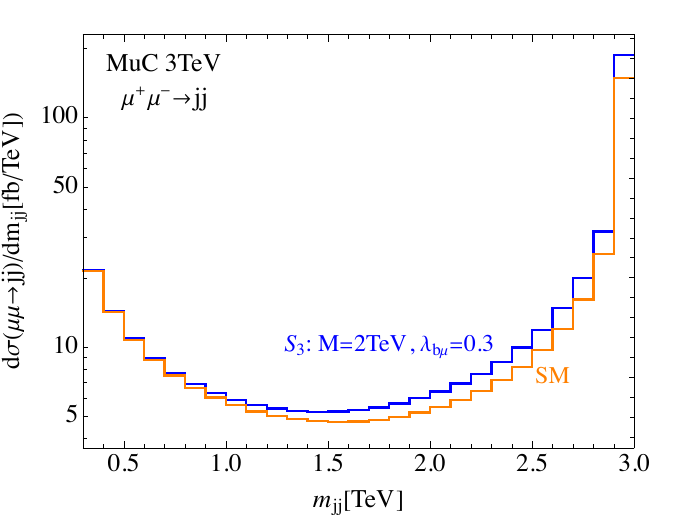}
\includegraphics[height=4.7cm]{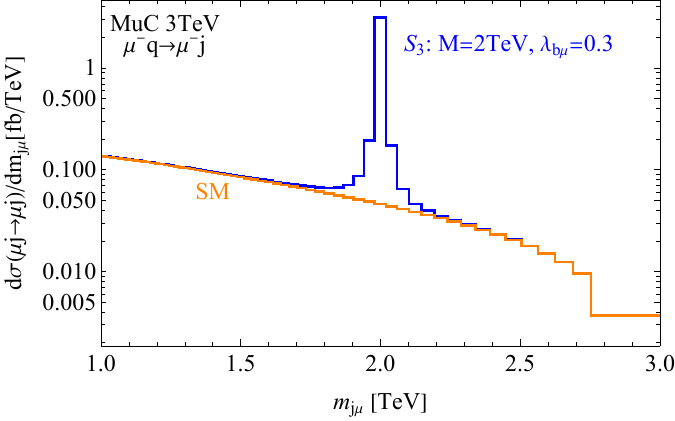}
\caption{\label{fig:S3XsecMuC} Differential cross sections for inclusive  $\mu^+ \mu^- \to j j$ (left) and $\mu^- q \to \mu^- j$ (right) for a 3 TeV MuC. The SM cross section is shown in orange, while the prediction for the $S_3$ resonance with $m_{S_3} = 2$~TeV and coupling $\lambda_{b\mu} = 0.3$ is shown in blue.}
\end{figure}

In case of leptoquark exchanged in the $t$-channel, since no sharp resonance or feature is present, as can be seen from the differential cross section in Fig.~\ref{fig:S3XsecMuC}\,(left), the dominant contribution to the sensitivity comes from the precise cross section measurement in the last bins. In Ref.~\cite{Qian:2021ihf} it was shown that studying the rapidity distribution of the final state jets could provide additional handles to increase the signal to background ratio for leptoquarks.

\subsection{Di-tau: {\boldmath$\mu \bar{\mu}$}$\to \tau^+ \tau^-$}

This process is sensitive to the $s$-channel exchange of a $Z^\prime$ vector boson coupled to both muon and tau leptons. It is particularly relevant for the baryophobic models such as the $U(1)_{L_\mu - L_\tau}$ detailed in Section~\ref{sec:ZprimeLmuLtau}.
A proper analysis of this channel is complicated by the fact that tau leptons decay into neutrinos, implying that the total invariant mass can not be {properly} reconstructed. Other kinematical variables, such as the transverse mass $m_T$, are therefore typically considered. These, however, can only be obtained by letting the tau leptons decay and performing a full-fledged collider analysis (see e.g. \cite{Faroughy:2016osc}). This is well beyond the scope of our paper. Instead, to estimate conservatively the reach in this channel we consider directly the di-tau invariant mass as our kinematical variable but, to take into account the reconstruction issues due to the neutrinos in the final state, we consider only very wide $m_{\tau\tau}$ bins: [0.15-0.5-2-3] TeV and [0.15-0.5-2-5-10] TeV for the MuC3 and MuC10, respectively. We checked that modifying these bins does not affect substantially the result, as expected in light of the result shown in Fig.~\ref{fig:ZprimeB3Lmu_mumujj} for $\mu^+\mu^- \to jj$. Furthermore, we assume conservatively the efficiency for the di-tau detection to be 70\%, see~\cite{Huang:2021nkl}.

\subsection{Di-muon: {\boldmath$\mu \bar{\mu}$}$\to \mu^+ \mu^-$}

The partonic-level process which by far dominates the di-muon production cross section is the Bhabha scattering $\mu^+ \mu^- \to \mu^+ \mu^-$, see Fig.~\ref{fig:FDiagrams}.
This signature is employed in the context of our $U(1)_{L_\mu - L_\tau}$ detailed in Section~\ref{sec:ZprimeLmuLtau}. 
The SM  differential cross section (orange) as well as a benchmark $Z^\prime$ contribution (blue) are shown in Fig.~\ref{fig:ZprimeXsecMuC} (right). We impose a cut on the lab-frame rapidity $|y_{\mu^\pm}| < 2$ by integrating numerically the triple differential cross section from Eq.~\eqref{eq:triplediffXsec}.
The di-muon invariant mass bins are constructed following the muon $p_T$ resolution described in App.~\ref{app:Detector}.

As in the IDY channel, the SM cross section first decreases and then increases again when approaching the collider energy. Similarly to the exchange of the SM neutral electroweak bosons, a $Z^\prime$ can be exchanged both in the $s$ and $t$ channels, inducing the typical interference pattern exhibited in Fig.~\ref{fig:ZprimeXsecMuC}. 
For masses $M < \sqrt{s_0}$, the sensitivity is obtained by a combination of the resonance peak and the precise measurement of the cross section in the last bin, similarly to what is seen in the di-jet channel.
We note that, if the $Z^\prime$ mass is sufficiently small ($M \ll \sqrt{s_0}$), the cross section measurement in the last bin depends only on the resonance coupling $g_{Z^\prime}$.
On the other hand, if $M \gg \sqrt{s_0}$ then the NP effect can be described as a four-muon contact interaction and the sensitivity is dominated by the precision measurement of the cross section in the highest invariant mass bins.

This final state, as well as the $\tau^+ \tau^-$ one, have been already studied in Refs.~\cite{Huang:2021nkl,Capdevilla:2021rwo,Capdevilla:2021kcf} in the context of a $U(1)_{L_\mu - L_\tau}$ model. However, the previous literature does not employ muon PDFs. Instead, the PDF effect is approximated by the real soft photon emission process $\mu^+ \mu^- \to \ell^+ \ell^- \gamma$, where the photon is not detected. On the other hand, the production process of a $Z^\prime$ in association with a hard photon improves the sensitivity for low masses and could offer additional handles to detect the resonance. Since, as we will see, $\mu \mu \to \mu\mu$ covers a large portions of the parameter space already at the 3\,TeV MuC, we will not consider the $Z^\prime \gamma$ processes in this work.

\subsection{Mono-lepton plus jet: {\boldmath$\mu \bar{\mu}$}$\to \mu^- j$}
\label{sec:resonant_production}

This channel offers the unique possibility of producing, in a $2 \to 2$ process, an $s$-channel resonance that couples directly to quarks and leptons, thanks to the quark content inside the muon, see the relevant diagrams in Fig.~\ref{fig:FDiagrams} for the partonic process $\mu q \to \mu q$. When the mass of a leptoquark is lower than $\sqrt{s_0}$ of the MuC, the dominant signal would appear as a peak in the invariant $\mu j$ mass distribution, $m_{\mu j}$. As an example, we show in Fig.~\ref{fig:S3XsecMuC} (right) the $S_3$ leptoquark resonance (blue) over the SM background for a 3 TeV MuC (orange). An analogous peak appears in the $U_1$ leptoquark $s$-channel exchange. For more details about the leptoquark models see Section~\ref{sec:LQ}. Due to the absence of additional hard leptons or jets in the final state, this process is different than the more typical on-shell single-production of leptoquarks (see e.g. Ref.~\cite{Qian:2021ihf}).

Notice that the PDF luminosities involving a muon and a quark are rapidly decreasing for higher $m_{\mu j}$ (see Fig.~\ref{fig:lumimumu}), because the quark content of the lepton vanishes in the limit $x \to 1$. As a result, the sensitivity to the resonance peak is stronger at values of $m_{\mu j}$ lower than the collider energy $\sqrt{s_0}$.

Lastly, if the mass of a NP mediator is larger than the collider energy, the induced contact interaction mostly manifests in the highest energy bins available, compatibly with the luminosity decrease. Yet, the sensitivity reach in this case is considerably weaker than in the case of the $t$-channel exchange in $\mu^+ \mu^- \to j j$, as it can be seen from the derived limits in Section~\ref{sec:LQ}.

\subsection{Leptoquark pair production: {\boldmath$\mu \bar{\mu}$}$\to {\rm LQ} \overline{{\rm LQ}}$}

This channel is dominated by the partonic process $\mu^+ \mu^- \to {\rm LQ}\, \overline{{\rm LQ}}$ for $M_{\rm LQ}$ close to $\sqrt{s_0}/2$, where all the PDFs except the muon one (and muon neutrino, to a lesser extent) are completely negligible. We thus compute analytically the partonic cross section considering the $Z/\gamma$ exchange in the $s$-channel, the quark exchange in $u$-channel and their interference, see the diagrams in Fig.~\ref{fig:FDiagrams}, and then convolute with muon PDFs to obtain the total cross section.
We consider only the case of on-shell leptoquarks. 
In order to estimate the MuC reach on this channel we requiring that at least 100 events of pair-produced leptoquarks are generated. {This number is compatible with the one obtained by the more detailed analysis of \cite{Asadi:2021gah,Qian:2021ihf}, assuming a $\Br(LQ \to b \mu) \sim \mathcal{O}(1)$.  As one might expect, we find that the leptoquark mass reach from pair production is approximately given by $M_{\rm LQ}^{\rm reach} \approx \sqrt{s_0}/2$, and this is fairly independent on the precise number of events required.}

A full collider simulation of this process and its backgrounds is beyond the purpose of this paper, and we refer to \cite{Asadi:2021gah,Qian:2021ihf} for a more detailed study of both leptoquark pair and single production (by fusion with a $Z/\gamma$ or via radiation off of a $b$ or $\mu$) at muon colliders.

\section{Signatures at a hadron collider}
\label{sec:hadron}

\begin{figure}[t]
\centering
\noindent
$q \bar q \to \mu^+ \mu^-$ \\
\includegraphics[height=3cm]{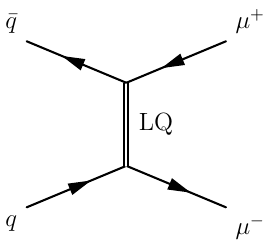} \quad \quad
\includegraphics[height=3cm]{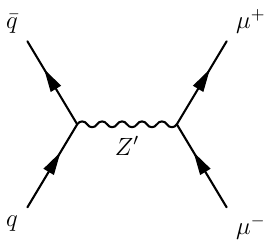} \\[0.4cm]
$q \bar q \to 4 \mu$ \quad \quad \quad \quad \quad \quad
\quad \quad \quad
$g g  \to {\rm LQ} \overline{{\rm LQ}}$ \\
\includegraphics[height=3cm]{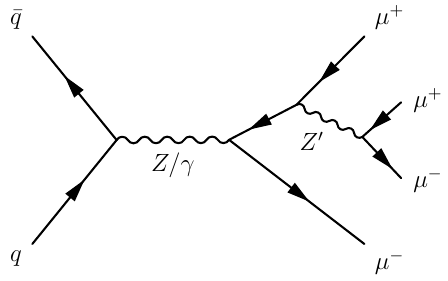} \quad \quad
\includegraphics[height=3cm]{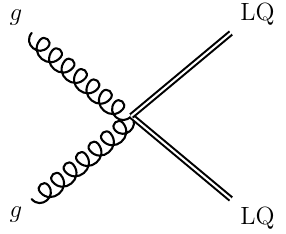} \\
\caption{\label{fig:FCCDiagrams} Sample Feynman diagrams for the partonic processes relevant at hadron colliders. For the scalar leptoquark $S_3$ one should exchange $q \leftrightarrow \bar{q}$.}
\end{figure}

In this Section, we highlight the processes at hadron colliders subject to our numerical studies. We use the current LHC data from CMS and ATLAS collaborations to set the $95\%$ CL limits that define the targeted parameter space for all considered models to be explored at future colliders.

\subsection{Di-muon: $p p  \to \mu^+ \mu^-$}

Following Ref.~\cite{Greljo:2017vvb}, a short-distance new physics above the electroweak scale contributing to the (semi)leptonic $B$-meson decays, generically predicts a correlated effect in the Drell–Yan (DY) process ($p p \to \mu^+ \mu^-$). This applies to all  tree-level mediators considered in this work. In particular, a $Z'$ would show up as an $s$-channel resonance, while a leptoquark would lead to a non-resonant effect via a $t$-channel contribution, see Fig.~\ref{fig:FCCDiagrams} for the respective Feynman diagrams. Should the mass of these mediators be above the accessible di-muon invariant mass spectrum, their impact would be described by a four-fermion quark-lepton interaction considered in Section~\ref{sec:contact}. Such interactions modify the high-invariant mass tails of the DY process~\cite{Cirigliano:2012ab, deBlas:2013qqa, Gonzalez-Alonso:2016etj, Faroughy:2016osc, Greljo:2017vvb, Cirigliano:2018dyk, Greljo:2018tzh,Bansal:2018eha,Angelescu:2020uug, Farina:2016rws, Alioli:2017nzr, Raj:2016aky, Schmaltz:2018nls, Brooijmans:2020yij,Ricci:2020xre,Fuentes-Martin:2020lea,Alioli:2017ces,Alioli:2017jdo,Alioli:2018ljm,Alioli:2020kez,Panico:2021vav,Sirunyan:2021khd,ATLAS:2021pvh,Marzocca:2020ueu,Afik:2019htr,Alves:2018krf,Greljo:2021kvv}. After specifying the quark flavour structure for a given operator, the sensitivity in the tails can be compared to those from the low-energy flavour physics. 

The production cross section depends crucially on the quark flavours involved in the initial state.  For example, quark-flavour universal $ Z' $ models with $B/L_\mu \sim \mathcal{O}(1)$ and MFV in the quark sector are already very well tested by current DY data at LHC. The dominant production channel in these models is due to the valance quarks, and it is enhanced because of their large PDFs. In this work, we only consider models in which the dominant couplings are with the heavy flavours and which can evade LHC searches thanks to the suppression from the sea quark PDFs. In Section~\ref{sec:ZprimeB3Lmu} we investigate the $U(1)_{B_3 - L_2}$ gauge extension of the SM where the $Z'$ primarily interacts with the third generation of quarks and second generation of leptons. The dominant DY channel in this model is the $b \bar b$ fusion. In Section~\ref{sec:LQ}, we derive the DY limits on the leptoquark models. While the main results are summarised in the aforementioned sections, the technical details of the numerical studies are discussed in Appendix~\ref{app:FCChh}.

\subsection{Multilepton: $p p  \to 4 \mu$}

The multilepton production at hadron colliders is relevant for a class of $Z'$ models in which the coupling to leptons is considerably larger than the coupling to quarks. The case study example is the $U(1)_{L_\mu - L_\tau}$ gauge model considered in Section~\ref{sec:ZprimeLmuLtau}. The Drell-Yan channels correlated with the $b \to s \mu^+ \mu^-$ decays:  $b \bar b$, $s  \bar s$, $s \bar b$ and $b \bar s$  are not only PDF-suppressed but are induced from a relatively small coupling in comparison with the muonic coupling. As we will show later, even the FCC-hh will have difficulties discovering such a scenario in $p p \to \mu^+ \mu^-$. On the other hand, a $Z'$ can be emitted from a muon in the charged or neutral current Drell-Yan leading to three or four muons in the final state after $Z'$ decays. A representative Feynman diagram for $p p \to \mu^+ \mu^- Z' \to 4 \mu$ is shown in Fig.~\ref{fig:FCCDiagrams}. In this work, we calculate the HL-LHC and the FCC-hh discovery projections (the technical details are left for the Appendix~\ref{app:FCChh}). A qualitative comparison is made with Ref.~\cite{Capdevilla:2021kcf} which derives the $95\%$ CL limits at the HL-LHC setting up a different analysis strategy.

\subsection{Leptoquark pair production: $ p p  \to {\rm LQ} \overline{{\rm LQ}} $}

Leptoquarks are colored particles and therefore easily produced in hadron colliders thanks to the QCD interactions~\cite{Blumlein:1996qp,Kramer:1997hh,Kramer:2004df,Diaz:2017lit,Dorsner:2018ynv,Borschensky:2020hot,Dorsner:2014axa,Borschensky:2021hbo}. The representative Feynman diagram is shown in Fig.~\ref{fig:FCCDiagrams}. The production cross section is set by $\alpha_s$ and by the leptoquark mass and its spin.\footnote{The cross section for the vector leptoquark $U_1$ also depends on another model-dependent coupling, see Eq.~\eqref{eq:U1gaugekappa}. For concreteness, we will assume that $U_1$ is a massive gauge boson of a Yang-Mills theory.}
When a leptoquark coupling to a quark and a lepton ($q\ell$-LQ) is large, there is an additional contribution to the pair production coming from the $t$-channel lepton exchange~\cite{Dorsner:2014axa}. This contribution is numerically relevant only in the parameter space where the Drell-Yan already provides better constraints. 

While the $q\ell$-LQ coupling can be neglected in the production of leptoquark pairs which is mainly due to QCD, it is very important for the decay channels. In our model examples, a leptoquark decay to $\mu j$ has a sizeable branching ratio, see Section~\ref{sec:LQ}. The FCC-hh projections for the scalar leptoquark pair production in $\mu^+ \mu^- j j $ final state have been derived in Ref.~\cite{Allanach:2019zfr}. We translate these bounds for the vector leptoquark using the toolbox of Ref.~\cite{Dorsner:2018ynv} based only on the total cross section and neglecting the differences in the kinematics. Other on-shell leptoquark production  mechanisms at hadron colliders, such as the single~\cite{Alves:2002tj,Hammett:2015sea,Mandal:2015vfa,Dorsner:2018ynv} and the resonant~ \cite{Ohnemus:1994xf,Buonocore:2020nai,Buonocore:2020erb,Greljo:2020tgv} production, are left for future studies. A common expectation is  that for leptoquarks dominantly coupled to heavy quarks, the phenomenology at hadron colliders is charted mainly by the pair production and the non-resonant contributions to the Drell-Yan, see~\cite{Dorsner:2018ynv,Schmaltz:2018nls,Buonocore:2020erb}.

\section{Contact interactions}
\label{sec:contact}

New physics states heavier than the energies accessible for on-shell production can still leave a trace in higher-dimensional operators of the SM effective field theory (SMEFT).
Discovering a new contact interaction, albeit at high energies, would still provide a valuable piece of information about the new physics. For example, measuring the contact interactions in the high-$p_T$ tails and establishing a correlation {with the anomalies in $B$-meson decays would exclude solutions to the anomalies due to light mediators, thus narrowing} down the set of possible ultraviolet completions. In particular, two effective operators in the SMEFT that match at tree-level to the low-energy operators relevant for $bs\mu\mu$ anomalies (and semileptonic decays in general) are
\beq
    \mathcal{L}_{\rm SMEFT} \supset [C_{\ell q}^{(1)}]_{22ij} (\bar L^2_L \gamma_\alpha L^2_L) ( \bar Q^i_L \gamma^\alpha Q^j_L ) + [C_{\ell q}^{(3)}]_{22ij} (\bar L^2_L \gamma_\alpha \sigma^a L^2_L) ( \bar Q^i_L \gamma^\alpha \sigma^a Q^j_L ) ~,
    \label{eq:SMEFT_ops}
\eeq
where $Q_L^i$ and $L_L^i$ are the SM left-handed quark and lepton weak doublets and $i,j=1,2,3$ are flavour indices. The flavour alignment is to the down-quark mass basis, $Q_L^i = ( V^*_{j i} u_L^j, d_L^i)^T$, $L_L^i = (\nu_L^i, e_L^i)^T$,
where $u_L^i$, $d_L^i$, $e_L^i$ fields are already the mass-eigenstates and the neutrinos are assumed to be massless.

At the LHC and the FCC-hh, these operators give a correction to the high invariant mass neutral-current DY tails $p p \to \mu^+ \mu^-$~\cite{Greljo:2017vvb} as well as to charged-current DY $p p \to \mu \nu$. For the latter we adapt the prospects derived in Ref.~\cite{Farina:2016rws}.\footnote{We translate $[C_{\ell q}^{(3)}]_{ijkl} = W / (2 v^2) \delta_{ij} \delta_{kl}$, where $W$ is the oblique EW parameter defined in Ref.~\cite{Barbieri:2004qk}, and rescale the bound by factor $1/\sqrt{2}$ to account for the fact that we have no contribution to the electron channel.}
At muon colliders these operators contribute to the high invariant mass di-jet production from both the neutral-current ($\mu^+ \mu^- \to j j $) and charged-current processes ($\mu^+ \nu_\mu \to j j  + h.c.$). The details of our numerical calculations (di-muon and di-jet resolution, PDFs, systematics, statistical treatment, etc.) are collected in the Appendices.
A further improvement in sensitivity, that we do not pursue in this work, could be obtained by asking one jet to be $b$-tagged when the dominant interaction involves bottom quarks, see~\cite{Afik:2019htr,Marzocca:2020ueu}. The high-energy tails are thus a complementary probe of new physics to mesonic decays and are useful to set new constraints on the flavour structure of new physics.

\begin{figure}[t]
\centering
\includegraphics[height=6.4cm]{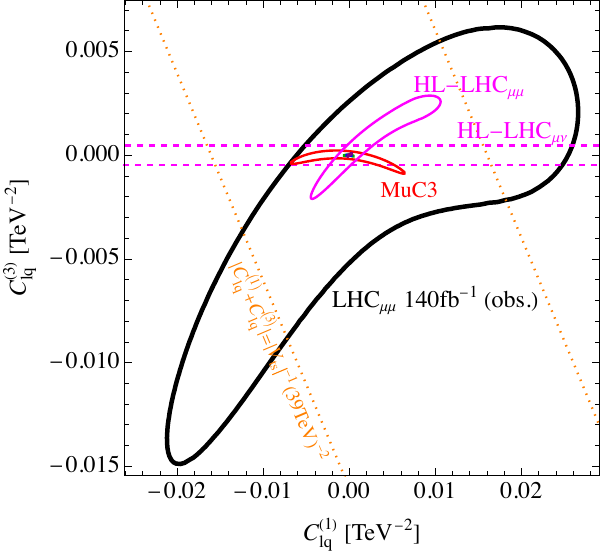} ~
\includegraphics[height=6.7cm]{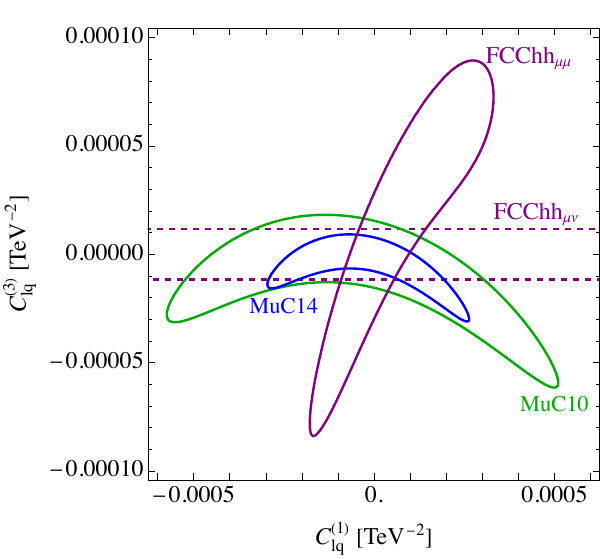}
\caption{\label{fig:EFTlimits_SU3} Sensitivity reach (95\% CL) for the quark flavour-universal scenario on the two EFT coefficients $C_{lq}^{(1)}$ and $C_{lq}^{(3)}$, in TeV$^{-2}$, for different colliders. The right plot is a zoomed-in version near the origin of the left plot.
For the future prospects at hadron colliders, we also include the sensitivity on $C_{lq}^{(3)}$ from the charged-current channel $p p \to \mu \nu$ (dashed lines), adapting the bound from Ref.~\cite{Farina:2016rws}.
The dotted orange lines are the solutions of $|C_{lq}^{(1)}+C_{lq}^{(3)}|= |V_{ts}^{-1}|\,(39 \rm TeV)^{-2}$ and illustrate an expectation from the $bs\mu\mu$ anomalies assuming MFV, see Eq.~\eqref{eq:size}.}
\end{figure}

\subsection{MFV scenario}

To begin with, we first study the MFV scenario in the quark sector. We assume the $U(3)_Q$ flavour symmetry in Eq.~\eqref{eq:SMEFT_ops}, leaving us with two universal and real parameters: $[C_{lq}^{(1)}]_{22ij} = C_{lq}^{(1)} \delta_{ij}$ and $[C_{lq}^{(3)}]_{22ij} = C_{lq}^{(3)} \delta_{ij}$. 
Breaking the symmetry by the insertions of the quark Yukawa matrices does not impact the Drell-Yan bound, however it induces contributions to mesonic decays~\cite{Greljo:2017vvb}.

In Fig.~\ref{fig:EFTlimits_SU3} we show the projected 95\% CL limits for various future colliders, compared with the present exclusion from the recast of the CMS search~\cite{CMS:2021ctt} (solid black line), see App.~\ref{app:FCChh} for details. Shown in the right plot is a zoom-in view around the origin of the left plot. 

Interestingly, MuC and FCC-hh probe complementary directions in the parameter space. While the MuC3 shows just slightly better sensitivity than the HL-LHC, the MuC10 is comparable with the FCC-hh. One of the reason for this is that at hadron colliders the production cross section is enhanced by the valence quarks. As we show in the next Section, when the dominant interaction is to heavy quark flavours, already MuC3 is comparable with the FCC-hh.

\subsection{Addressing $bs\mu\mu$ anomalies}

Matching Eq.~\eqref{eq:SMEFT_ops} to the low-energy EFT at tree-level gives the relevant operator controlling the $b \to s \mu^+ \mu^-$ decays,
\be
    C_{sb\mu\mu} = \left( [C_{\ell q}^{(1)}]_{2223} + [C_{\ell q}^{(3)}]_{2223} \right)~.
\ee
In realistic models the $sb\mu\mu$ interaction is rarely generated alone; it comes along with the flavour-diagonal interactions, such as $bb\mu\mu$. In motivated flavour scenarios that aim at addressing the flavour puzzle and providing sufficient protection for approximate accidental symmetries, such as the $U(2)^3$ flavour symmetry in the quark sector~\cite{Barbieri:2011ci}, the $bb\mu\mu$ contact interaction is expected to be enhanced with respect to the $sb\mu\mu$. 
{For this reason, in the following we consider two scenarios: only $sb\mu\mu$ contact interaction and only the $bb\mu\mu$ one (where $sb\mu\mu$ is assumed to be $\sim |V_{ts}|$ suppressed with respect to the flavour diagonal one and thus negligible).}
Assuming the $C_{\ell q}^{(1)} = C_{\ell q}^{(3)}$ alignment, we study high-energy constraints on the following effective Lagrangian:
\begin{equation}
	\mathcal{L}_{\rm EFT} = C_{bb\mu\mu} \; (\bar{b}_L \gamma_\alpha b_L)(\bar \mu_L \gamma^\alpha \mu_L) + \left( C_{sb\mu\mu} \; (\bar{s}_L \gamma_\alpha b_L)(\bar \mu_L \gamma^\alpha \mu_L)  + \text{h.c.} \right)~.
	\label{eq:EFT_collider}
\end{equation}

\begin{figure}[t]
\centering
\includegraphics[height=8cm]{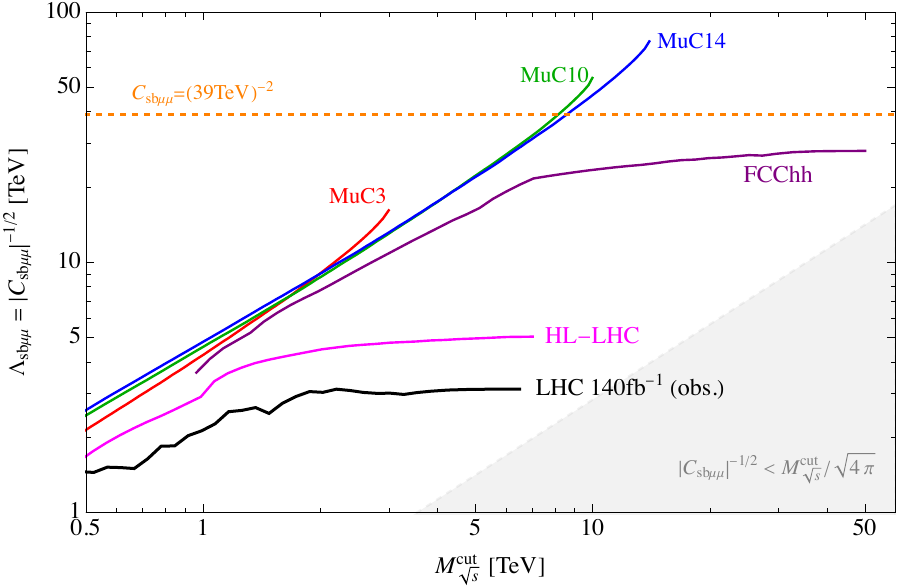}
\caption{\label{fig:EFTlimits_bs} Sensitivity reach (95\%CL) for the $(\bar{s}_L \gamma_\alpha b_L) (\bar{\mu}_L \gamma^\alpha \mu_L)$ contact interaction as function of the upper cut on the final-state invariant mass, compared to the value required to fit $bs\mu\mu$ anomalies (dashed orange line).}
\end{figure}

\begin{figure}[t]
\centering
\includegraphics[height=8cm]{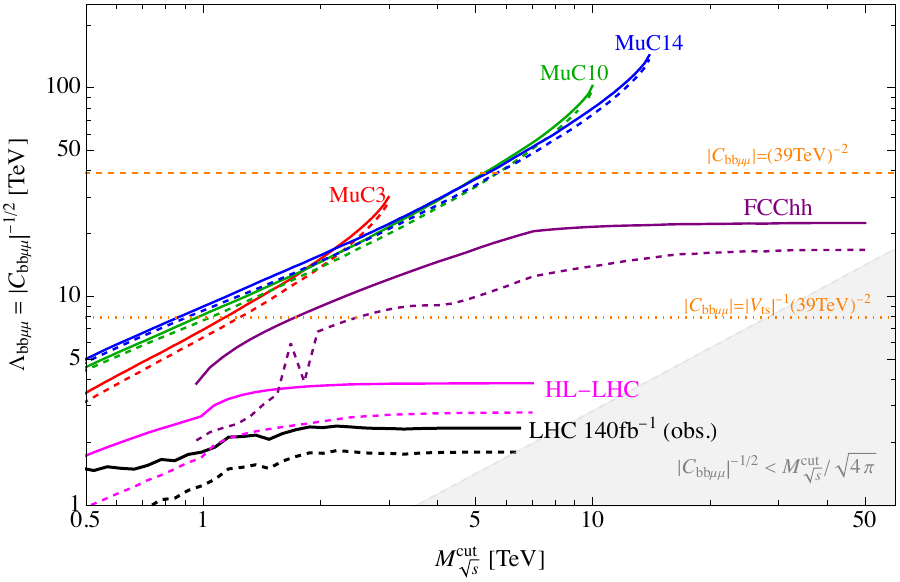}
\caption{\label{fig:EFTlimits_bb} Sensitivity reach (95\%CL) for the $(\bar{b}_L \gamma_\alpha b_L) (\bar{\mu}_L \gamma^\alpha \mu_L)$ contact interaction as function of the upper cut on the final-state invariant mass. Solid (dashed) lines represent the limit for positive (negative) values of $C_{bb\mu\mu}$. The orange dotted and dashed lines shows reference values in relation to the $bs\mu\mu$ anomalies fit, with or without a $1/V_{ts}$ enhancement of the $bb$ operator compared to the $bs$ one, respectively.}
\end{figure}

This choice of contact interactions has recently gained attention due to the LHCb anomalies. What LHCb reported so far are discrepancies in rare semileptonic $B$ meson decays with the underlying quark level transition $b \to s \mu^+ \mu^-$. The anomalous observables include: branching ratios~\cite{LHCb:2020zud,LHCb:2021awg,LHCb:2021vsc,LHCb:2014cxe,LHCb:2015wdu,LHCb:2016ykl,LHCb:2021zwz};
angular distributions~\cite{LHCb:2020lmf,LHCb:2020gog}; and the theoretically very clean~\cite{Hiller:2003js,Bordone:2016gaq,Isidori:2020acz} LFU ratios, \RK~\cite{LHCb:2017avl,LHCb:2021trn}. When interpreting all of the data in a low-energy effective field theory, a consistent picture of NP emerges (for recent global fits see~\cite{Altmannshofer:2021qrr,Geng:2021nhg,Alguero:2021anc,Hurth:2021nsi,Ciuchini:2020gvn,Li:2021toq,Kriewald:2021hfc,Cornella:2021sby}). The global significance of the NP hypothesis, including the look-elsewhere effect, was conservatively estimated to be $4.3\sigma$~\cite{Isidori:2021vtc}. Considering a single non-zero Wilson coefficient, the global fits identify two preferred scenarios: $\Delta C_9^\mu = -0.73 \pm 0.15$, or $\Delta C_9^\mu = - \Delta C_{10}^\mu = -0.39 \pm 0.07$~\cite{Altmannshofer:2021qrr}.\footnote{These are defined from the following effective Hamiltonian at the $m_b$ scale~\cite{Altmannshofer:2021qrr}:
\be
    \mathcal{H}_{\rm eff}^{\rm NP} \supset - \frac{4 G_F}{\sqrt{2}} V_{tb} V_{ts}^* \frac{\alpha}{4\pi} \left[ \Delta C_9^\mu (\bar s_L \gamma_\alpha b_L)(\mu \gamma^\alpha \mu) + \Delta C_{10}^\mu (\bar s_L \gamma_\alpha b_L)(\mu \gamma^\alpha \gamma_5 \mu) \right] + \text{h.c.}~. \nonumber
\ee
}
The latter can be conveniently translated in terms of the new physics effective operator $C_{sb\mu\mu}$. The best-fit point from~\cite{Altmannshofer:2021qrr} corresponds to
\beq\label{eq:size}
\left. C_{sb\mu\mu} \right|_{\rm best-fit} \approx (39\, \TeV)^{-2}~.
\eeq

The other possibility advocated by the global fit of semileptonic $B$-decays is the operator with the vectorial muon current, $\mathcal{O}_9^\mu \propto (\bar b_L \gamma^\mu s_L) (\bar \mu \gamma^\mu \mu)$, see e.g.~\cite{Altmannshofer:2021qrr}. At high energies (in the massless limit), the amplitudes with different muon chiralities do not interfere. We do not expect a big difference between the two cases and consider only the left-handed muon current. The effective operator with the vectorial muon current is instead explicitly realized in the two $Z^\prime$ models studied in the next section. The EFT limit is recovered when the $Z'$ is much heavier than the collider energy.

In Figs.~\ref{fig:EFTlimits_bs} and \ref{fig:EFTlimits_bb} we show the expected 95\% CL sensitivity on the EFT coefficients in Eq.~\eqref{eq:EFT_collider} for the future colliders listed in Table~\ref{tab:colliders}, as well as the observed bound from the recast of the CMS search~\cite{CMS:2021ctt}. The constraints are shown as lower limits on the effective scale $\Lambda_X = |C_X|^{-1/2}$ in TeV, as a function of the upper cut on the invariant mass of the final state, $M^{\rm cut}_{\sqrt{s}}$. We recall that in order for the EFT to be valid, the NP scale should be higher than the maximal experimental energy, in this case $M^{\rm cut}_{\sqrt{s}}$, hence: $M_{\rm NP} > M^{\rm cut}_{\sqrt{s}}$. Parametrising the EFT coefficients as $C = g_*^2 / M_{\rm NP}^2$, where $g_*$ describes the interaction strength between the NP and the SM, the loosest EFT validity constraint is derived for strongly coupled theories, $g_*^2 \sim 4 \pi$, from which we get the bound $|C|^{-1/2} > M^{\rm cut}_{\sqrt{s}} / \sqrt{4\pi}$.  The region that does not satisfy this condition is shaded in gray. This approximates the perturbative unitarity bound on $M_{\rm NP}$ for a given value of the coefficient $C$ \cite{DiLuzio:2017chi}.

As shown in Figs.~\ref{fig:EFTlimits_bs} and \ref{fig:EFTlimits_bb}, the FCC-hh comes close to the minimal scenario $C_{s b \mu\mu} = (39 \,{\rm TeV})^{-2}$ and can easily cover the more realistic scenario, $C_{b b \mu\mu} = \pm |V_{ts}|^{-1} (39\, \TeV)^{-2}$.\footnote{We neglect the RGE effects for the purpose of this comparison since the chosen operators do not renormalize under QCD~\cite{Alonso:2013hga,Celis:2017doq}.} The interference with the SM is negligible in the former case, while in the latter case it dominates the limit. The solid and the dashed lines in Fig.~\ref{fig:EFTlimits_bb} stand for the positive and the negative sign of $C_{b b \mu\mu}$, respectively. Finally, our projections agree with the previous studies for the HL-LHC~\cite{Greljo:2017vvb} and for the FCC-hh~\cite{Garland:2021ghw}. The latter reference also considers the effect of subleading backgrounds.

Similarly to the FCC-hh, the 3 TeV  MuC also comes close to the minimal scenario and can easily cover the more realistic scenario, as already anticipated for the case involving heavy flavours and suppressed production at hadron colliders (see also Ref.~\cite{Huang:2021biu} for similar results).
Both MuC10 and MuC14, instead, are able to completely test both scenarios.

At the MuC, the EFT limits for the highest invariant mass cut $=\sqrt{s_0}$, can be easily estimated by looking just at the partonic cross section $\hat\sigma(\mu^+ \mu^- \to j j)(m_{\mu\mu})$. For energies $m_{\mu\mu} \gg m_Z$ we have:
\begin{eqnarray}
    &&\hat\sigma(\mu^+ \mu^- \to j j)(m_{\mu\mu}) =  \nonumber \\
    &&\approx \frac{N_c}{48 \pi m_{\mu\mu}^2} \left(  \sum_{q_X} \sum_{Y=L,R} \left|g_Z^{q_X} g_Z^{\mu_Y} - e^2 Q^{q_X} + m_{\mu\mu}^2 C_{q_X q_X \mu\mu}|^2 + 2 m_{\mu\mu}^4 |C_{sb\mu\mu}\right|^2 \right) \approx \nonumber \\
    && \approx \frac{624\, \text{fb}}{(m_{\mu\mu} / \TeV)^2} \left( 1 + 2.35 C_{bb\mu\mu} m_{\mu\mu}^2 + 12.4 C_{bb\mu\mu}^2 m_{\mu\mu}^4 + 24.8 |C_{sb\mu\mu}|^2 m_{\mu\mu}^4 \right) ,
    \label{eq:EFT_Mcdi-jet_analytic}
\end{eqnarray}
up to relative corrections of $\mathcal{O}\left(m_Z^2 / m_{\mu\mu}^2\right)$, where the sum over $q_X$ runs on all quarks with both left and right chiralities, except for the top.
For instance, assuming 1 ab$^{-1}$ of luminosity and a 2\% systematic uncertainty, for the 3 TeV muon collider one gets a 95\%~CL bound $|C_{sb\mu\mu}| \lesssim (15\, \TeV)^{-2}$, very similar to the one shown in Fig.~\ref{fig:EFTlimits_bs} following from a full numerical study detailed in the Appendices.
In case of the flavour-diagonal $C_{b b\mu\mu}$ contact interaction in Fig.~\ref{fig:EFTlimits_bb}, given the energy and precision of the measurements, the MuC limits are dominated by the interference with the SM, as can be quickly derived from Eq.~\eqref{eq:EFT_Mcdi-jet_analytic}, hence the limits are essentially symmetric between positive and negative values.

\section{$Z'$ models}
\label{sec:Zprime}

Heavy neutral vectors are obvious candidates for mediating $b \to s \mu \mu$ transitions at the tree-level~\cite{Buras:2013qja,Altmannshofer:2014cfa,Greljo:2015mma,Crivellin:2015mga,Celis:2015ara,Falkowski:2015zwa,Crivellin:2016ejn,Boucenna:2016qad,Ko:2017lzd,Alonso:2017uky,Alonso:2017bff,Bhatia:2017tgo,Bian:2017xzg,King:2018fcg,Bonilla:2017lsq,Calibbi:2019lvs,Ellis:2017nrp,Allanach:2018lvl,Altmannshofer:2019xda,Allanach:2020kss,Allanach:2021gmj,Davighi:2021oel,Bause:2021prv}.
The couplings most relevant for our discussion are the ones to the muon current, $g_{\mu\mu}$, to flavour-conserving quark currents, $g_{qq}$, and to the flavour-violating $sb$ current, $g_{sb}$.
The contribution to $b \to s \mu \mu$ transitions is proportional to the product $g_{sb} g_{\mu\mu}$, while the $B_s - \overline{B}_s$ mixing is proportional to $g_{sb}^2$, thus $g_{sb} \ll g_{\mu\mu} $.
The effect of flavour diagonal couplings $g_{qq}$ is typically the most relevant in the high-$p_T$ processes, due to the interference with the SM amplitude.
In the case of a hadron collider, it is particularly important to identify the size of $g_{qq}$ for each quark flavour $q$, since couplings to the first generation induce larger signals due to the PDF enhancement of valence quarks. There is no such distinction at a MuC in the inclusive di-jet searches, however, employing $b$-tagging would enhance the sensitivity to $g_{bb}$ with respect to other light flavours.

The quark flavour-universal case with $g_{qq} \sim g_{\mu\mu}$ was explored in Ref.~\cite{Greljo:2017vvb}, which concluded that the current LHC data already provide stringent constraints. Therefore, we expect future hadron colliders to perform better compared to MuCs in such scenarios. In light of this, we consider models in which the dominant quark coupling is to heavy flavours. There are two qualitatively different scenarios that still need to be studied.
Either $g_{sb} \ll g_{bb} \sim g_{\mu\mu}$, in which case the flavour-conserving couplings to quarks and to muons are of the same order and the flavour symmetry protects against excessive flavour violation, or $g_{sb} \sim g_{bb} \ll g_{\mu\mu}$, in which case all couplings to quarks are suppressed with respect to couplings to leptons. These two setups predict different phenomenologies and are therefore worth studying separately. The first scenario is naturally realised, for instance, by gauging $X = B_3 - L_\mu$ (Section~\ref{sec:ZprimeB3Lmu}).
The second scenario instead can be obtained by gauging $X = L_\mu - L_\tau$ (Section~\ref{sec:ZprimeLmuLtau}).

In the following we consider the two models separately. In both cases, the $Z^\prime$ coupling to $sb$ can be generated, for instance, via quark mixing with some vectorlike fermions after spontaneous breaking of the $U(1)_X$ gauge symmetry. In each scenario, we first carry out sensitivity studies at future colliders when such mixing is negligible and then we fix the mixing in order to fit the present $bs\mu\mu$ anomalies and perform a more focused study.

\subsection{$U(1)_{B_3 - L_\mu}$ model}
\label{sec:ZprimeB3Lmu}

Let us consider an extension of the SM gauge symmetry where the anomaly-free charge $X = B_3 - L_\mu$ is gauged.\footnote{The set of SM chiral fermions is minimally extended with three right-handed neutrinos which can be motivated by the smallness of the neutrino masses through a seesaw mechanism. One of them carries the same $X$ charge as $\mu_R$ as required by the chiral anomaly cancellation conditions.} Similar models have already been proposed as a way to address the $b s \mu\mu$ anomalies in Refs.~\cite{Alonso:2017uky,Bonilla:2017lsq,Allanach:2020kss}, to which we refer for more details. In the unbroken phase, the $U(1)_{B_3 - L_\mu}$ gauge boson $Z^\prime$ has a vectorial coupling to third-generation quarks and second-generations leptons.
A small coupling to the second-generation quark doublet is induced after spontaneous
symmetry breaking with a scalar field $\phi$, charged only under  $U(1)_{B_3 - L_\mu}$. The gauge-invariant operators 
$(\phi^\dagger D_\mu \phi) (\bar Q_L^2 \gamma^\mu Q_L^3)$ and $ \bar Q_L^2 H \phi b_R$ get generated after integrating out, 
for example, heavy vectorlike quarks. In particular, the latter operator is anyhow required by the CKM elements $V_{td}$ and $V_{ts}$ which are absent in the 
renormalisable model with the minimal matter content. The smallness of the 1-3 and 2-3 mixing in the quark sector is 
explained by the higher-dimensional operator breaking the accidental flavour 
symmetry of the renormalisable Lagrangian. In addition, the same operator indirectly 
induces the $Z' s b$ coupling in the 
broken phase after the rotation to the mass basis of the left-handed down quarks 
by a small angle $\theta_{sb}$. Thus, the model naturally predicts an approximate $U(2)^3$ flavour symmetry allowing for a 
TeV-scale new physics compatible with flavour bounds~\cite{Barbieri:2011ci}.

Assuming only the rotations for left-handed fermions and $\theta_{sb} \ll 1$, the leading $Z^\prime$ couplings to SM fermions are
\beq\begin{split} \label{eq:lagB3L2}
	\L_{Z^\prime_{B_3 - L_\mu}}^{\rm int} = & - g_{Z^\prime} Z^\prime_\alpha \left[ \frac{1}{3} \bar{Q}^3_L \gamma^\alpha Q^3_L + \frac{1}{3} \bar{b}_R \gamma^\alpha b_R + \frac{1}{3} \bar{t}_R \gamma^\alpha t_R -  \bar{L}^2_L \gamma^\alpha L^2_L - \bar{\mu}_R \gamma^\alpha \mu_R + \right. \\
	& \left. +  \left( \frac{1}{3} \epsilon_{sb}  \bar{Q}^2_L \gamma^\alpha Q^3_L  + \text{h.c.}  \right ) + \mathcal{O}(\epsilon_{sb}^2)\right]  ~,
\end{split}\eeq
where for convenience we introduced $\epsilon_{sb} \equiv \frac{1}{2} \sin 2\theta_{sb}$. Thus, the total decay width to the SM fermions for the $Z^\prime$ is
\be
    \Gamma_{Z^\prime_{B_3 - L_\mu}} \approx \frac{M_{Z^\prime} g_{Z^\prime}^2}{24 \pi} \left[3 + \frac{1}{3} \left( 4 + 4 |\epsilon_{sb}|^2 \right) \right]~,
    \label{eq:width_Zp_B3Lmu}
\ee
where the top mass is neglected (in the numerical study we keep finite $m_t$ effects). We consider $\Gamma_{Z^\prime}/M_{Z^\prime} < 0.25$ as a qualitative perturbativity criterion for the model.\footnote{By the optical theorem, the decay width of the resonance is related to the imaginary part of the two-point function starting at the one-loop level. Therefore, when the width becomes \emph{of the order} of the mass, the loop corrections are comparable with the tree level, which indicates a loss of perturbativity.} We neglect the muon-flavoured right-handed neutrino assuming $m_N > m_{Z'}/2$. Otherwise, it would contribute to the total $Z'$ decay width slightly changing the numerical results.

The measurement of neutrino trident production cross section sets an upper limit on the coupling to muons as a function of the mass~\cite{Altmannshofer:2014cfa},
\be
    g_{Z^\prime} < 2.0 \, \frac{M_{Z^\prime}}{\TeV} ~.
    \label{eq:nuTrident}
\ee

\begin{figure}[t]
\centering
\includegraphics[width=13cm]{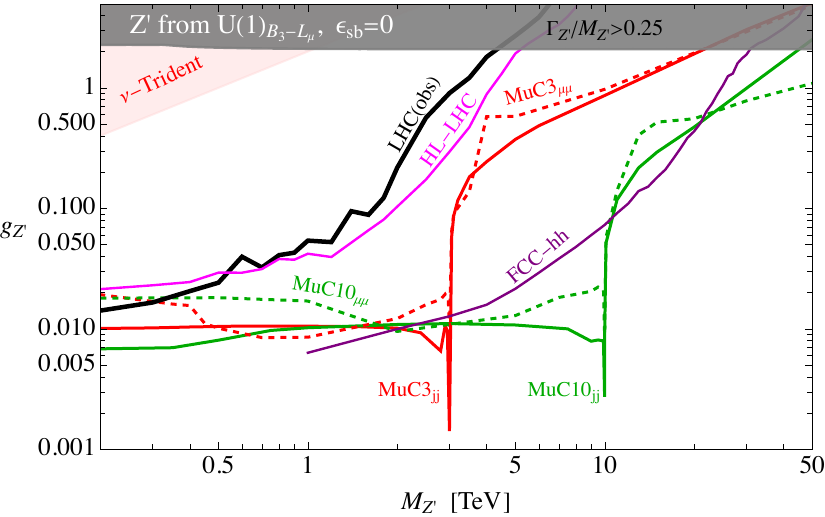}\\
\caption{\label{fig:Zp_B3-Lmu_bounds_noRK} Discovery reach at 5$\sigma$ for the $B_3 - L_\mu$ model with $\epsilon_{sb} = 0$, for different final states at each collider (as indicated by the labels). The region excluded at 95\%\,CL by LHC \cite{CMS:2021ctt} is above the black line while in the dark gray region the $Z^\prime$ has a large width, signaling a loss of perturbativity.}
\end{figure}

\subsubsection*{No-mixing scenario}

By assuming $\epsilon_{sb} = 0$ in Eq.~\eqref{eq:lagB3L2} we study the discovery potential of future machines for a $Z^\prime$ coupled only to muons (and muon neutrinos) and third generation quarks. We consider the signatures discussed in Section~\ref{sec:MuC} (for a MuC) and in Section~\ref{sec:hadron} (for a hadron collider).
Shown in Fig.~\ref{fig:Zp_B3-Lmu_bounds_noRK} are the present 95\%\,CL exclusion bounds from the recast of the CMS Drell-Yan analysis~\cite{CMS:2021ctt} (thick black line) and neutrino trident production (red region), as well as the 5$\sigma$ discovery prospects for various future colliders. The dark gray region corresponds to $\Gamma_{Z^\prime}/M_{Z^\prime} > 0.25$, {which signals a loss of perturbativity.}

We observe that MuC3 improves substantially from the HL-LHC prospects for all masses and MuC10 provides the best potential sensitivity (MuC14 would further improve on this). In this model FCC shows a sensitivity comparable, albeit somewhat weaker, than MuC10.

\begin{figure}[t]
\centering
\includegraphics[width=13cm]{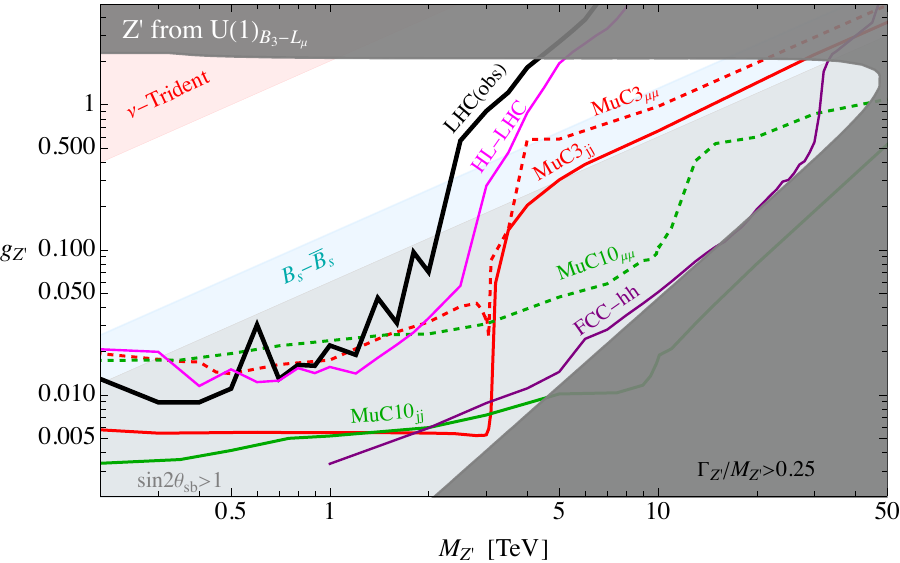}\\
\includegraphics[width=6.5cm]{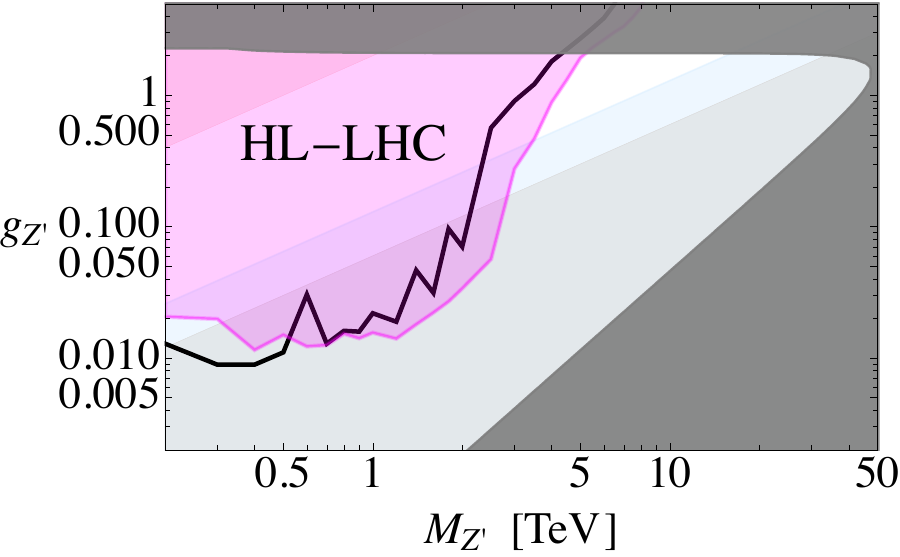} 
\includegraphics[width=6.5cm]{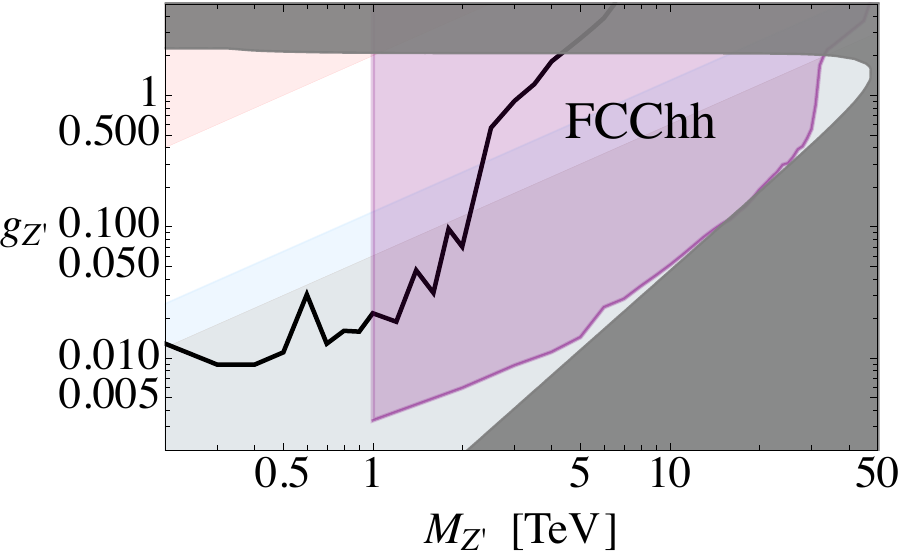} \\
\includegraphics[width=6.5cm]{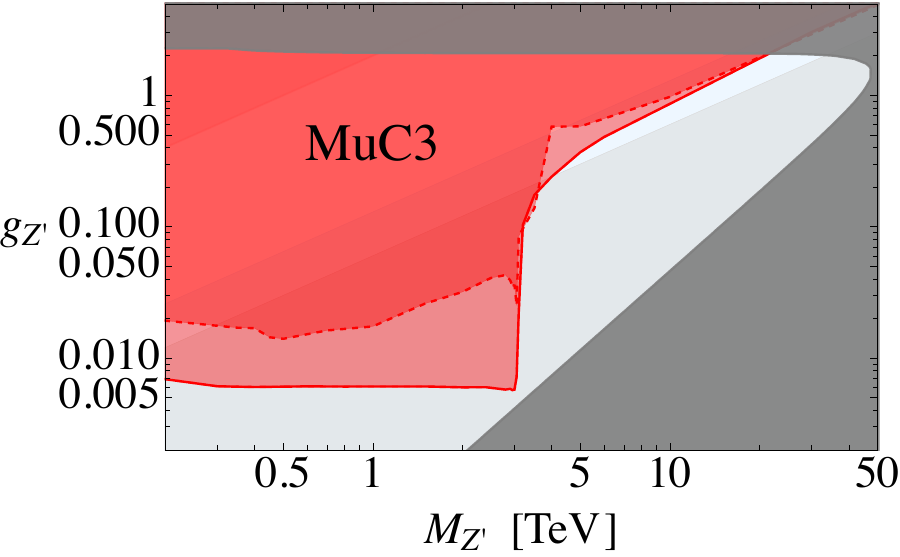}
\includegraphics[width=6.5cm]{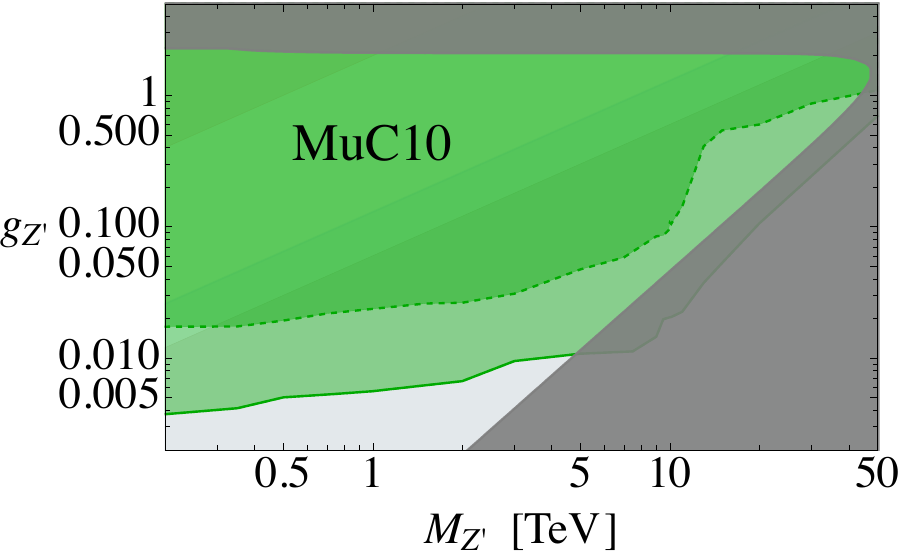}
\caption{\label{fig:Zp_B3-Lmu_bounds} Discovery reach at 5$\sigma$ for the $B_3 - L_\mu$ model. The fit to $bs\mu\mu$ anomalies is imposed everywhere, Eq. \eqref{eq:ZpB3Lmu_thetasb}. The region excluded at 95\%\,CL by LHC \cite{CMS:2021ctt} is above the black line, while the one excluded by  $B_s$ mixing is colored in light blue. The light gray region cannot provide a successful fit to $b\to s\mu\mu$ anomalies for values of $\sin 2\theta_{sb} < 1$, Eq.~\eqref{eq:ZpB3Lmu_thetasb}, while in the dark gray region the $Z^\prime$ has a large width, signaling a loss of perturbativity. The discoverable region at future colliders is the one on the side of the line where the corresponding label has been drawn. The smaller figures below the main figure highlight a single future collider at a time.}
\end{figure}

\subsubsection*{Addressing $bs\mu\mu$ anomalies}

Now we turn to a more specific study for $bs\mu\mu$ anomalies. For given values of $g_{Z^\prime}$ and $M_{Z^\prime}$, the mixing parameter required to fit the $bs\mu\mu$ anomalies is
\beq
	\epsilon_{sb} = - 1.7 \times 10^{-3} \left( \frac{M_{Z^\prime}}{g_{Z^\prime} \TeV} \right)^2 \left( \frac{\Delta C_9^\mu}{-0.73} \right) ~.
	\label{eq:ZpB3Lmu_thetasb}
\eeq
The same coupling ($Z'sb$) induces $B_s-\overline{B}_s$ mixing:
\beq
    C^1_{B_s} = \frac{\left(g_{Z^\prime} \frac{1}{3} \epsilon_{sb} \right)^2}{M_{Z^\prime}^2}~,
\eeq
which is constrained by $|C^1_{B_s}| < |V_{tb} V_{ts}^*|^2 / (9.2 \, \TeV)^2$~\cite{UTfit:2007eik,Silvestrini:2018dos}, implying a lower value for $g_{Z^\prime} \geq 0.125 \, \Delta C_9^\mu / (-0.73)  (M_{Z^\prime} / \TeV)$, if Eq.~\eqref{eq:ZpB3Lmu_thetasb} is imposed.
This also ensures that the condition $\theta_{sb} \ll 1$ is always satisfied in the acceptable region.
In this model the contributions to $D^0$-mixing is suppressed by $(\epsilon_{sb} V_{cs} V_{ub})^2$ or $(\epsilon_{sb}^2 V_{cs} V_{us})^2$, making the corresponding bound much less stringent than the $B_s$-mixing one.

We impose Eq.~\eqref{eq:ZpB3Lmu_thetasb} assuming the best fit value for $\Delta C_9^\mu$ in order to address the $bs\mu\mu$ anomalies in every point of the $(M_{Z'},g_{Z'})$ plane shown in Fig.~\ref{fig:Zp_B3-Lmu_bounds}.
To help the reader understand the projections for different future colliders, we report the same plot in the bottom part of the figure, considering only a single collider and coloring the potentially discoverable region.
The difference with the previous result is that, for a given mass, decreasing $g_{Z^\prime}$ requires larger values of $\epsilon_{sb}$ (i.e. $\theta_{sb}$).
The $B_s - \overline{B}_s$ constraint is shown as a light-blue region. The light gray region corresponds to values of masses and coupling that would require $\sin 2\theta_{sb} > 1$ in order to fit the $b s \mu \mu$ anomalies.\footnote{While this is not possible in the model of Ref.~\cite{Allanach:2020kss}, it could in principle be achieved by having some vectorlike quarks with large charges. However, it is likely that in this case couplings to $b$ and $t$ quarks also receive $\mathcal{O}(1)$ corrections from the mixing. In any case, this region is excluded by $B_s$-mixing.}

Our recast of the present CMS Drell-Yan search excludes at 95\%\,CL the region above the thick black line, which includes all viable couplings in the mass range $200 \,\GeV < M_{Z^\prime} \lesssim 2 \,\TeV$ (we do not study the model for masses below 200 GeV).
Instead, for the future colliders listed in Table~\ref{tab:colliders}, we show the 5$\sigma$ discovery reach, where the region in parameter space above the corresponding line is discoverable. Regarding hadron colliders, the HL-LHC projected bounds will only be slightly improved, while the FCC-hh will be able to completely test this scenario via resonance searches in $p p \to \mu^+ \mu^-$.
The 3 TeV MuC is also able to completely cover the viable parameter space not already excluded by the existing constraints. For masses below 3 TeV, a resonance could be observed directly in both di-muon (dashed red) and di-jet channels (solid red), while a heavier $Z^\prime$ would show-up as a smooth deviation from the SM in the highest invariant mass bins. A 10 TeV MuC would further improve the sensitivity allowing for resonance searches even for a heavier $Z'$, up to the region allowed by perturbativity and meson mixing.

\subsection{$U(1)_{L_\mu - L_\tau}$ model}
\label{sec:ZprimeLmuLtau}

We next consider a heavy $Z^\prime$ arising from a spontaneously broken $U(1)_{L_\mu - L_\tau}$ gauge symmetry.\footnote{The chiral anomaly cancellation conditions require introducing right-handed neutrinos if one wants a muonic vector current as in Section~\ref{sec:ZprimeB3Lmu}.} As shown in Ref.~\cite{Altmannshofer:2014cfa}, this model is a viable candidate to address the $bs\mu\mu$ anomalies. In the limit of a vanishing kinetic mixing with the hypercharge ($X_{\mu\nu} B^{\mu\nu}$), the $Z'$ does not couple to quarks at the renormalisable level when only the minimal matter content is present.\footnote{The kinetic mixing is typically induced by RGE when additional fields charged under $L_\mu-L_\tau$ are present, see Appendix A.3 of Ref.~\cite{Greljo:2021npi}. In that case, one gets a loop-suppressed $p p \to Z'$ from valence quarks that can be relevant (see Eq.~(3.10) in Ref.~\cite{Greljo:2022dwn}). However, this contribution can be removed by a small tree-level kinetic mixing.}
For instance, those couplings can be generated, after the spontaneous symmetry breaking, via mixing with heavy vectorlike quarks charged under $L_\mu - L_\tau$. Therefore, the quark couplings (including $b$ quarks) are expected to be much smaller than couplings to muons and taus. Let $\epsilon_b$ and $\epsilon_s$ be some small mixings with vectorlike quarks of the corresponding left-handed quark doublets in the down-quark mass basis of third and second generation, respectively. The relevant $SU(2)_L$ invariant  $Z^\prime$ interactions are
\beq
\begin{split}\label{eq:ZpMuTa}
	\L_{Z^\prime_{L_\mu - L_\tau}}^{\rm int} =& - g_{Z^\prime} Z^\prime_\alpha \left[ \bar{L}^2_L \gamma^\alpha L^2_L + \bar{\mu}_R \gamma^\alpha \mu_R - \bar{L}^3_L \gamma^\alpha L^3_L - \bar{\tau}_R \gamma^\alpha \tau_R + \right. \\
	& \left. + |\epsilon_b|^2 \bar{Q}^3_L \gamma^\alpha Q^3_L
	+ |\epsilon_s|^2 \bar{Q}^2_L \gamma^\alpha Q^2_L + \left( \epsilon_b \epsilon_s^*  \bar{Q}^2_L \gamma^\alpha Q^3_L  + {\rm h.c.} \right) + \ldots\right] ~.
\end{split}
\eeq
The total decay width of the $Z^\prime$ is
\be
\begin{split}
    \Gamma_{Z^\prime_{L_\mu-L_\tau}} \approx \frac{M_{Z^\prime} g_{Z^\prime}^2}{24 \pi} \left[6 + 3 \left( 2 |\epsilon_s|^4 + 4 |\epsilon_s|^2 |\epsilon_b|^2 + 2 |\epsilon_b|^4 \right) \right]~,
    \label{eq:width_Zp_LmuLtau}
\end{split}
\ee
where the top mass is neglected (in the numerical study we keep the physical $m_t$). Similarly to Section~\ref{sec:ZprimeB3Lmu}, we impose $\Gamma/M < 0.25$ as the perturbativity limit and neglect the right-handed neutrinos in $Z'$ decays. The constraint on $g_{Z^\prime}$ from neutrino trident production is the same as in Eq.~\eqref{eq:nuTrident}.

\begin{figure}[t]
\centering
\includegraphics[width=13cm]{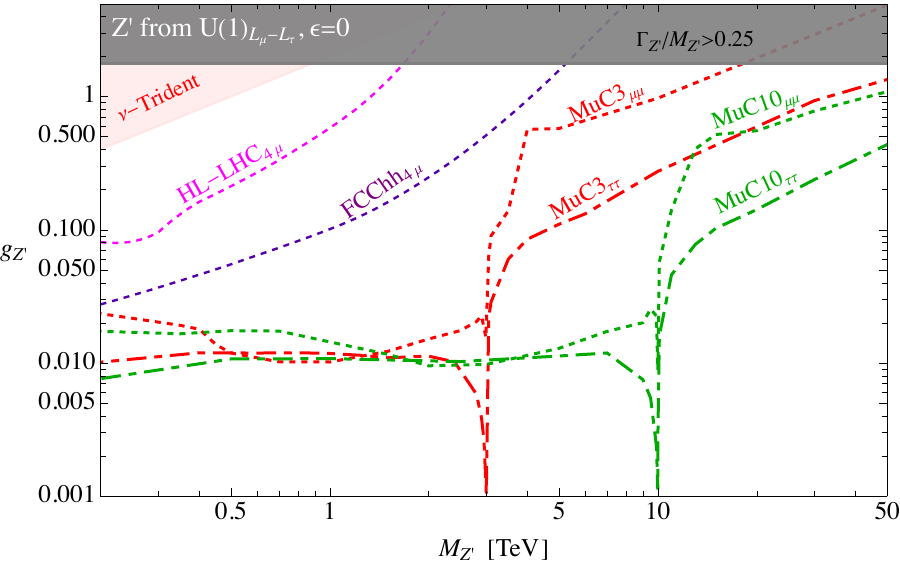}\\
\caption{\label{fig:Zp_Lmu-Ltau_bounds_noRK} Discovery reach at 5$\sigma$ for the $L_\mu - L_\tau$ model with $\epsilon_{s} = \epsilon_{b} = 0$ in Eq.~\eqref{eq:ZpMuTa}. In the dark gray region the $Z^\prime$ has a large width, signaling a loss of perturbativity.}
\end{figure}

\subsubsection*{Quark-phobic $Z^\prime$}

We first focus on a scenario where the $Z^\prime$ is quark-phobic ($\epsilon_s = \epsilon_b = 0$) and derive the present 95\%\,CL exclusion bounds as well as the future discovery projections. The results are reported in Fig.~\ref{fig:Zp_Lmu-Ltau_bounds_noRK}, where we show the 5$\sigma$ sensitivity reach for various future colliders.
The shaded regions are analogous to the ones from the previous Section.

This case (not surprisingly) illustrates a situation in which even the MuC3 outperforms the FCC-hh (since the $Z'$ is both quark-phobic and leptophilic). Our limits for MuC3 agree well with those obtained in Ref.~\cite{Huang:2021nkl}.
Other channels at a MuC such as: $\mu^+ \mu^- \to \ell^+ \ell^- \gamma$ (where $\ell = \mu, \tau$), $\mu^+ \mu^- \to \nu \bar{\nu} \gamma$, and $\mu^+ \mu^- \to \gamma Z^\prime$ offer additional handles to pinpoint the properties of the $Z^\prime$ boson, see Refs.~\cite{Huang:2021nkl,Capdevilla:2021rwo,Capdevilla:2021kcf}.

\begin{figure}[t]
\centering
\includegraphics[width=13cm]{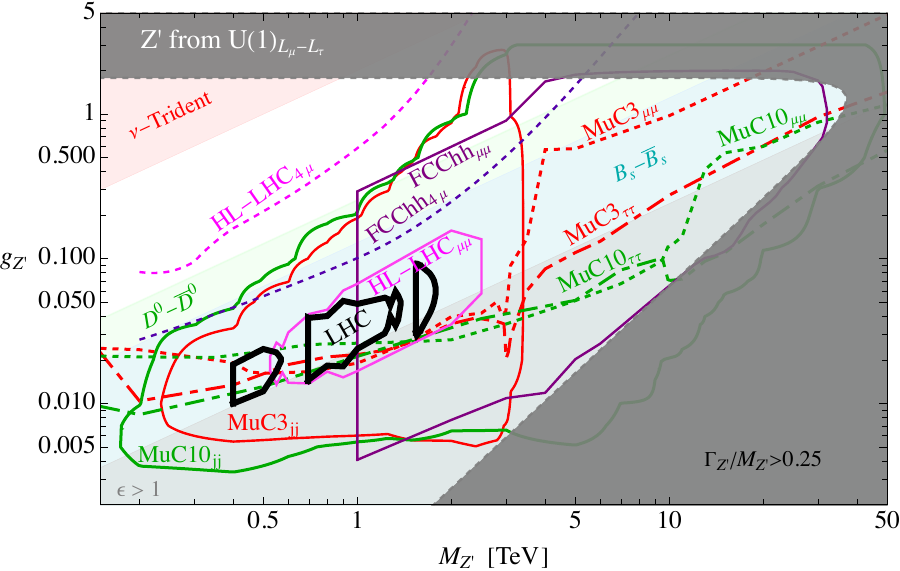}\\
\includegraphics[width=6.5cm]{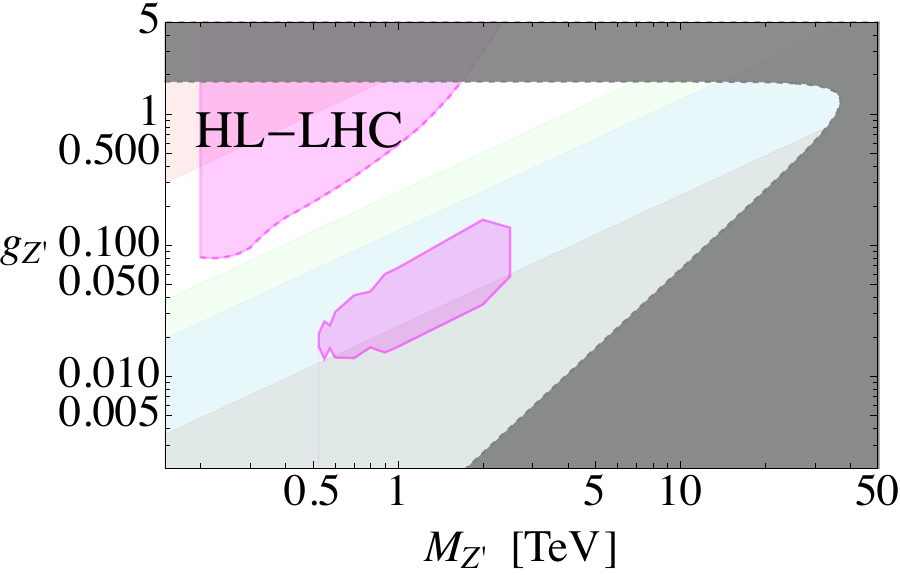}
\includegraphics[width=6.5cm]{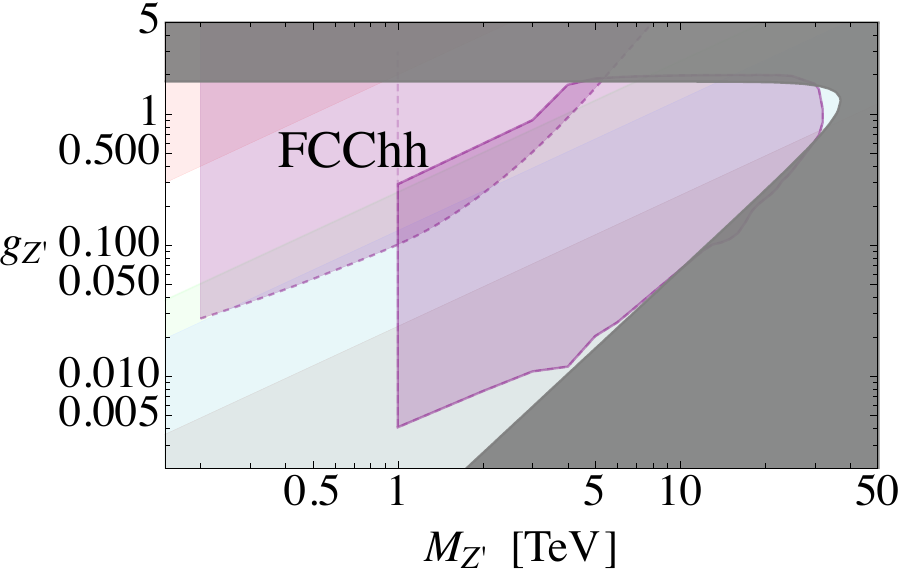} \\
\includegraphics[width=6.5cm]{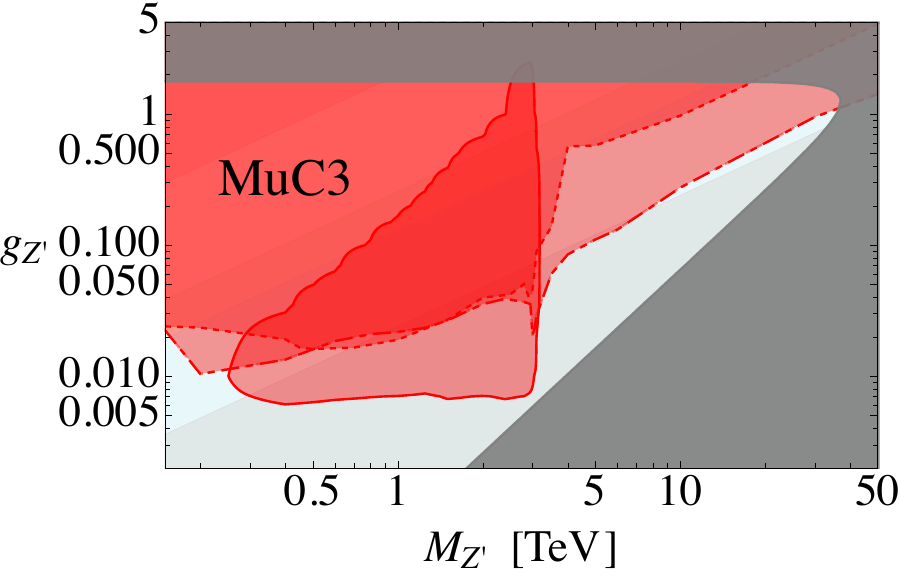}
\includegraphics[width=6.5cm]{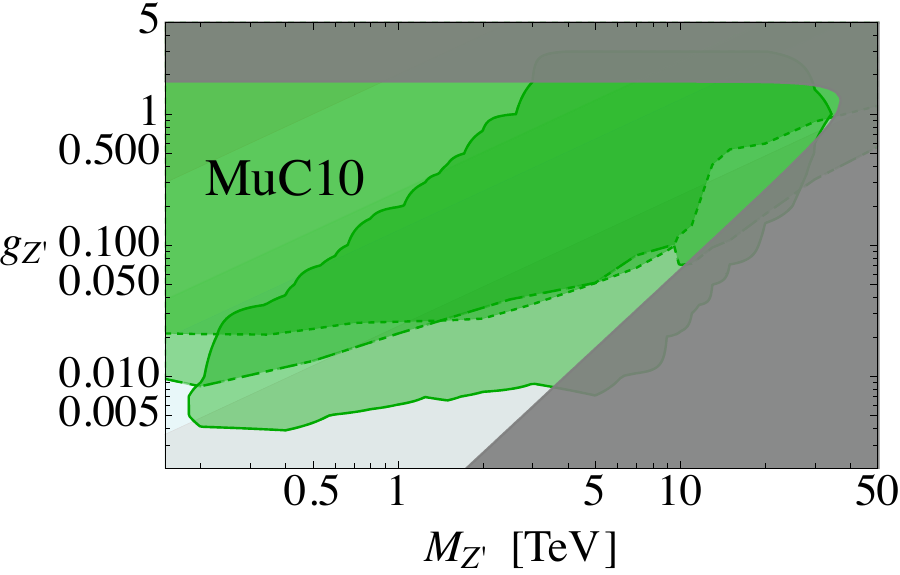}
\caption{\label{fig:Zp_Lmu-Ltau_bounds} Discovery reach at 5$\sigma$ for the $L_\mu - L_\tau$ model. The fit to $bs\mu\mu$ anomalies is imposed, Eq.~\eqref{eq:ZpLmuLtau_epsRK}. The regions excluded by $B_s$ and $D^0$ mixings and neutrino trident production are colored in light blue, green and red, respectively. The light gray region requires $\epsilon > 1$ to fit $b\to s\mu\mu$ anomalies, while the dark gray region has $\Gamma_{Z^\prime} / M_{Z^\prime} > 0.25$. Our recast of the present LHC search~\cite{CMS:2021ctt} excludes at 95\%\,CL the region \emph{inside} the thick black lines. The discoverable region at future colliders is the one on the side of the line where the corresponding label has been drawn. The smaller figures below the main figure highlight a single future collider at a time.}
\end{figure}

\subsubsection*{Addressing $bs\mu\mu$ anomalies}

In order to address the $bs\mu\mu$ anomalies, the product of the mixing parameters is set to:
\beq
	\epsilon_b \epsilon_s^* = - 5.7 \times 10^{-4} \left( \frac{M_{Z^\prime}}{g_{Z^\prime} \TeV} \right)^2 \left( \frac{\Delta C_9^\mu}{-0.73} \right) ~.
	\label{eq:ZpLmuLtau_epsRK}
\eeq
Even after imposing $\Delta C_9^\mu = -0.73$, we are left with other free parameters besides $M_{Z'}$ and $g_{Z'}$. Our goal here is to study the case where $| \epsilon_s / \epsilon_b | \sim \mathcal{O}(1)$  which is qualitatively different from the model in Section~\ref{sec:ZprimeB3Lmu}. For concreteness, in our numerical analysis we assume $\epsilon_b = - \epsilon_s$ and $\Im \epsilon_b = 0$. With this simplification, we are able to plot our results in the $(M_{Z'},g_{Z'})$ plane.

Analogously to Section~\ref{sec:ZprimeB3Lmu},  the $B_s$ mixing,  $C^1_{B_s} = - (g_{Z^\prime} \epsilon_s^* \epsilon_b)^2 / M_{Z^\prime}^2$, together with Eq.~\eqref{eq:ZpLmuLtau_epsRK}, imply the lower limit $g_{Z^\prime} > 0.125 M_{Z^\prime}/\TeV$. The $D^0-\overline{D}^0$ mixing gives another constraint on the parameters: $C^1_{D^0} = (g_{Z^\prime} V^*_{us} V_{cs} |\epsilon_s|^2)^2 / M_{Z^\prime}^2 < 2.5 \times 10^{-13} \GeV^{-2}$ \cite{UTfit:2007eik,Silvestrini:2018dos}, corresponding to $g_{Z^\prime} > 0.25 M_{Z^\prime}/\TeV$. Interestingly, $D^0$ mixing provides stronger constraints than $B_s$-mixing in this model.

Our main results are shown in Fig.~\ref{fig:Zp_Lmu-Ltau_bounds}. The present CMS $pp \to \mu^+\mu^-$ data~\cite{CMS:2021ctt} exclude at 95\%\,CL the region \emph{inside} the thick black lines. For the future colliders listed in Table~\ref{tab:colliders}, the parameter space discoverable at 5$\sigma$ is the one \emph{on the side} of the corresponding line where the label is shown. To help the reader better understand the sensitivity reach for each collider, below the main plot in Fig.~\ref{fig:Zp_Lmu-Ltau_bounds} we report four smaller plots where the 5$\sigma$ discover sensitivity for each collider is isolated and shaded.
Note that, in the case of $p p \to \mu^+\mu^-$ at hadron colliders or $\mu^+\mu^- \to jj$ at MuCs, the only accessible region is for intermediate values of $g_{Z^\prime}$. According to Eq.~\eqref{eq:ZpLmuLtau_epsRK}, for a given $Z^\prime$ mass the couplings to quarks are inversely proportional to $g_{Z^\prime}$.  Since too large $g_{Z^\prime}$ values imply too small couplings to quarks, and vice versa, there is always a suppression in $\sigma \times \mathcal{B}$ for the two processes. The di-muon searches at FCC-hh and the di-jet searches at MuC can cover a much larger parameter space than the one accessible at (HL-)LHC but are still unable to cover the viable parameter space for the $bs\mu\mu$ anomalies. In this respect, the most optimal channels at MuCs are $\mu \mu \to \mu\mu$ and $\mu\mu \to \tau \tau$, that even at a 3 TeV MuC are enough to completely cover the leftover parameter space. At hadron colliders, the most promising channel is $p p \to 4 \mu$. Let us emphasize that even the HL-LHC can make significant progress, while the FCC-hh would fully cover the viable parameter space.

\section{Leptoquark models}
    \label{sec:LQ}

Leptoquarks~\cite{Dorsner:2016wpm} are hypothetical particles that can couple quarks to leptons at the renormalizable level. They are motivated by the idea of quark-lepton unification hinted at by the hypercharge quantization in the SM. Leptoquarks are also the only other mediators, in addition to colorless vectors, that generate the semileptonic effective operators in Eq.~\eqref{eq:SMEFT_ops} at the tree level. Interesting for our discussion are the scalar $S_3$, with the SM quantum numbers $({\bf \bar{3}},{\bf 3},1/3)$, and the vector $U_1 \sim ({\bf 3},{\bf 1},2/3)$.\footnote{We do not consider $U_3 \sim ({\bf 3},{\bf 3},2/3)$ since its phenomenology is partially covered by the $U_1$ case. Similarly, we did not consider a colorless vector triplet in Section~\ref{sec:Zprime}. The $SU(2)_L$ gauge symmetry will in both cases predict additional correlated signatures.} Both are a viable single-mediator solution of the $bs\mu\mu$ anomalies~\cite{Angelescu:2021lln}.

In this Section, we investigate the discovery prospects at future colliders for the $S_3$ and $U_1$ leptoquarks. We extend the SM minimally with a single heavy field (ignoring the UV origin of its mass) and focus on the renormalisable interactions with the left-handed SM fermions. We consider two different cases regarding the flavour structure of such interactions. First, we assume an exact $U(2)_{Q_L}$ quark-flavour symmetry under which the first two generations $Q_L^i$ ($i=1,2$) form a doublet, while the third-generation $Q_L^3$ is a singlet. In addition, we assume an exact $U(1)_{\mu - LQ}$ symmetry under which $L_L^2$ and the leptoquark are oppositely charged. This can be achieved by gauging one out of many possible anomaly-free lepton flavour non-universal $U(1)$ extensions of the SM, see~\cite{Greljo:2021npi}.  In this case, the only allowed coupling will be to $Q_L^3$ and $L_L^2$.
In the second scenario, we aim at addressing the $ b s \mu \mu$ anomalies by minimally adding a direct leptoquark coupling to $Q_L^2$.

Relaxing our assumptions, it is conceivable to formulate scenarios with dominant couplings to taus or even to new exotic fermions consistent with the low-energy flavour bounds and proton decay. A famous example is the $U(2)_{L}$ flavour structure in the leptonic sector, advocated for a combined explanation of the $b s \mu \mu$ anomalies and $R_{D^{(*)}}$, see e.g. Ref.~\cite{Buttazzo:2017ixm}. These scenarios would require a different strategy since $LQ \to \mu j$ would be a subdominant decay mode. In addition, the interesting leptoquark mass range would also be more restricted by the perturbative unitarity, implying lighter states. For the future prospects on leptoquarks decaying to third generation leptons see~\cite{Cerri:2018ypt,CidVidal:2018eel}. In what follows, we analyze the minimal scenarios where such additional structures are neglected.

\subsection{Scalar leptoquark $S_3$}
\label{sec:LQscalar}

We start with the leptoquark $S_3 \sim ({\bf \bar{3}},{\bf 3},1/3)$~\cite{Dorsner:2016wpm}.
The interaction Lagrangian reads
\begin{eqnarray}
{\cal L}_{S_3}^{\rm int} = \lambda_{i\mu} \, \overline{Q_L^{i \,c}} \,\epsilon\,  \sigma^I L_L^2 S_3^I + \text{h.c.}~,
\label{eq:S3LQcoup}
\end{eqnarray}
where $\epsilon=i\sigma _2$. We assume a real coupling matrix for simplicity. The leptoquark triplet can be written as
\begin{eqnarray}
\left( S_3^I \sigma^I \right)\equiv\left(
\begin{array}{cc}
 S_3^{(1/3)}    & \sqrt 2 S_3^{(4/3)}  \\
\sqrt 2 S_3^{(-2/3)}     & -S_3^{(1/3)}
\end{array}\right)~,
\label{eq:S3_def}
\end{eqnarray}
where the superscript indicates the electric charge of each $S_3$ component. We assume a degenerate mass spectrum for the components, as expected from the $SU(2)_L$ gauge symmetry.
In the mass basis of SM fermions, the interaction Lagrangian \eqref{eq:S3LQcoup} becomes
\be
 {\cal L}_{S_3}^{\rm int} = - \lambda_{i\mu} S_3^{(1/3)} (V_{ji}^* \overline{u^{j \, c}_L} \mu_L + \overline{d^{i \, c}_L} \nu_\mu) + \sqrt{2} \lambda_{i\mu} \left( V_{ji}^* S_3^{(-2/3)} \overline{u^{j \, c}_L} \nu_\mu - S_3^{(4/3)} \overline{d^{i \, c}_L} \mu_L \right) + \text{h.c.}~.
\ee
The total decay width of $S_3$, in the limit of vanishing fermion masses, is given by
\begin{eqnarray}
\Gamma_{S_3} = \frac{|\lambda_{b\mu}|^2 + |\lambda_{s\mu}|^2}{8\pi} M_{S_3}~,
\end{eqnarray}
assuming only $\lambda_{b\mu}$ ($i=3$) and $\lambda_{s\mu}$ ($i=2$) different from zero.
The perturbativity limit $\Gamma_{S_3}/M_{S_3}<0.25$ is considered, as previously.

\begin{figure}[t]
\centering
\includegraphics[width=13cm]{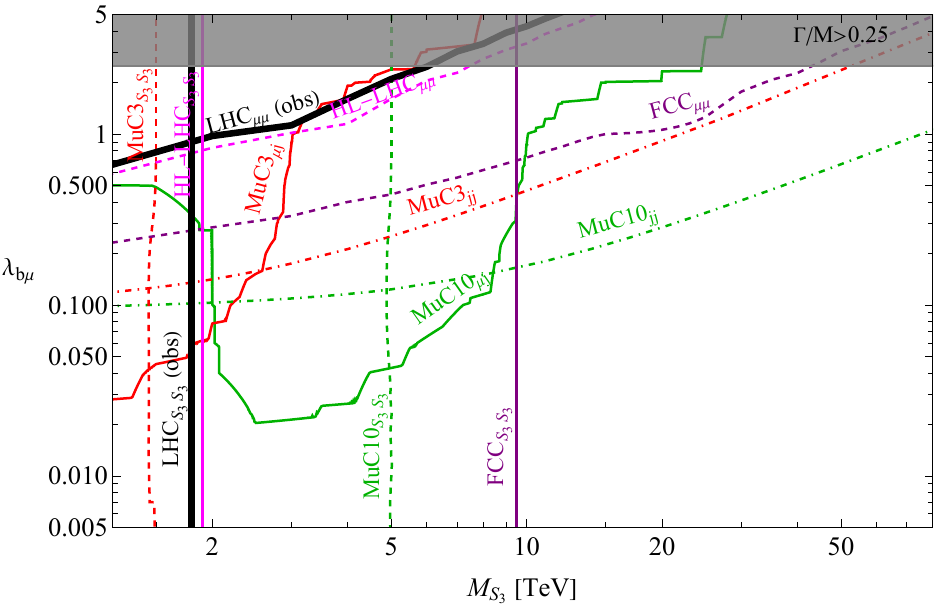}
\caption{\label{fig:S3_bounds_bmu}
The 5$\sigma$ discovery prospects at future colliders for the $S_3$ leptoquark assuming the $U(2)^3$ quark flavour symmetry and the exclusive leptoquark coupling to muons (see Section~\ref{sec:LQscalar}). The present LHC exclusions at 95\%\,CL are shown as a thick black line. The perturbativity limit $\Gamma_{S_3}/M_{S_3}<0.25$ is violated in the grey region. The labels for various colliders and processes are on the discoverable side of a curve.}
\end{figure}

\subsubsection*{$U(2)^3$ symmetric case}

Imposing an unbroken $U(2)^3$ quark flavour symmetry, and assuming $S_3$ to be charged under the muon number, only the $\lambda_{b\mu}$ coupling is allowed.
This symmetry is broken in the SM by light quark masses and by the mixing of third-generation quarks with the first two via the CKM matrix. This is an approximate symmetry of the SM Yukawa sector, where the largest symmetry-breaking term is $|V_{ts}|\approx 0.04$. Assuming the minimal $U(2)^3$ breaking and no breaking of $U(1)_\mu$ as in the SM, the expected sizes of other non-zero leptoquark couplings are $|\lambda_{s\mu}| \sim |V_{ts} \lambda_{b\mu}|$ and $|\lambda_{d\mu}| \sim |V_{td} \lambda_{b\mu}|$, see Refs.~\cite{Bordone:2017anc,Buttazzo:2017ixm,Fuentes-Martin:2019mun,Marzocca:2021miv}. Those can be neglected in our collider study.

In Fig.~\ref{fig:S3_bounds_bmu} we show the present 95\%\,CL limits from LHC searches (thick black) and the $5\sigma$ discovery prospects for future colliders (various colored lines), considering only $\lambda_{b\mu} \neq 0$, as motivated by the aforementioned approximate flavour symmetry of the SM. The leptoquark pair production at the LHC sets a robust lower limit on the mass even for small couplings, while the Drell-Yan process excludes a region with the large coupling even for higher masses. Interestingly, the HL-LHC $5\sigma$ discovery region is only marginally larger than the present 95\%\,CL exclusions. Nevertheless, the FCC-hh will drastically improve both the Drell-Yan and the leptoquark pair production reach.  

Regarding muon colliders, we find that a 3\,TeV MuC would have a comparable reach from the IDY process as the FCC-hh from the DY, while the MuC10 would easily surpass the FCC-hh. On the other hand, the FCC-hh provides a far superior prospects on pair-production, being able to discover on-shell leptoquarks with masses of almost 10 TeV, compared to only 5\,TeV for the MuC10. The resonant leptoquark production at the MuC10 could probe a unique region in the parameter space compared with other production mechanisms at muon colliders. However, this region can easily be covered at the FCC-hh.

\begin{figure}[t]
\centering
\includegraphics[width=13cm]{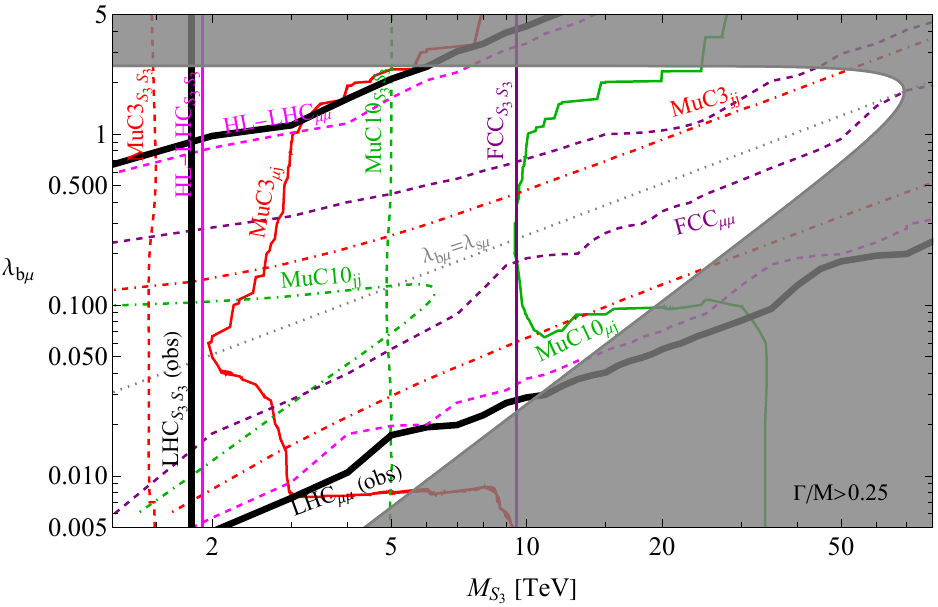} \\
\includegraphics[width=6.5cm]{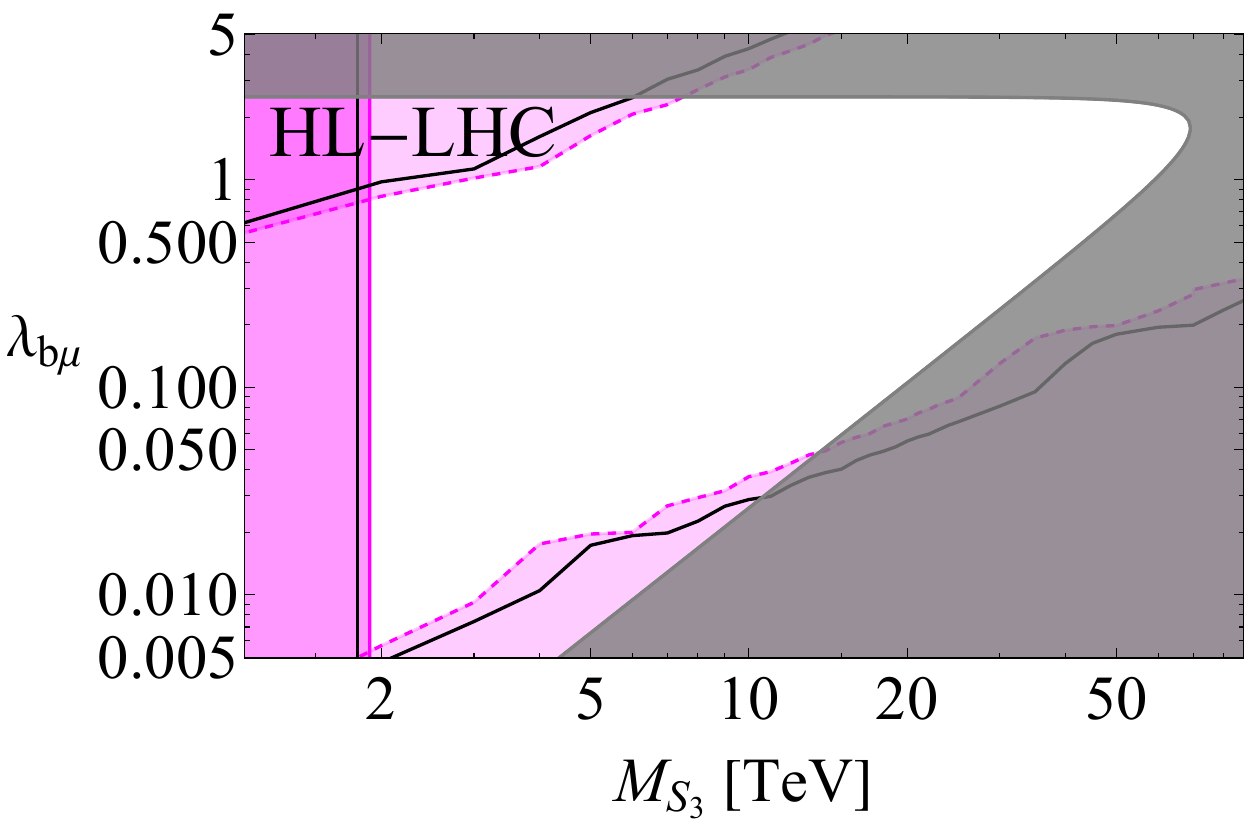}
\includegraphics[width=6.5cm]{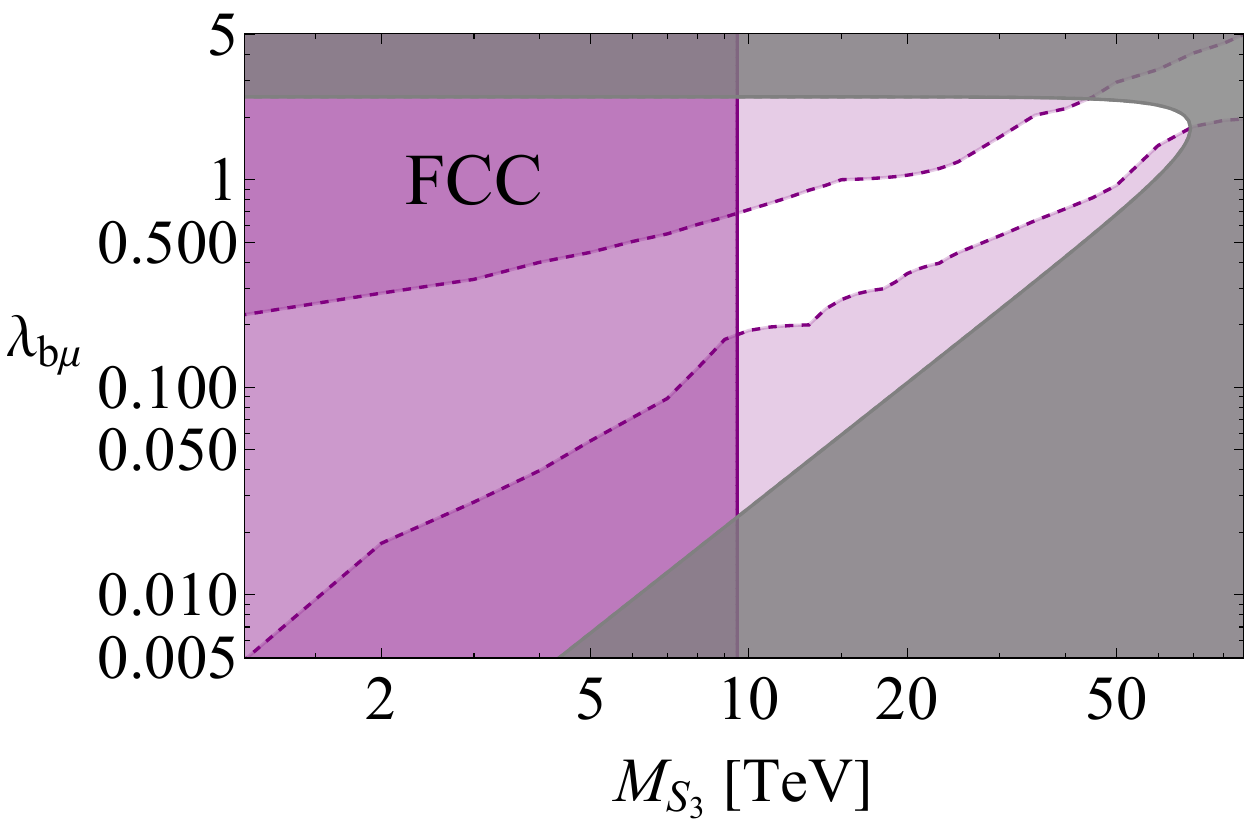} \\
\includegraphics[width=6.5cm]{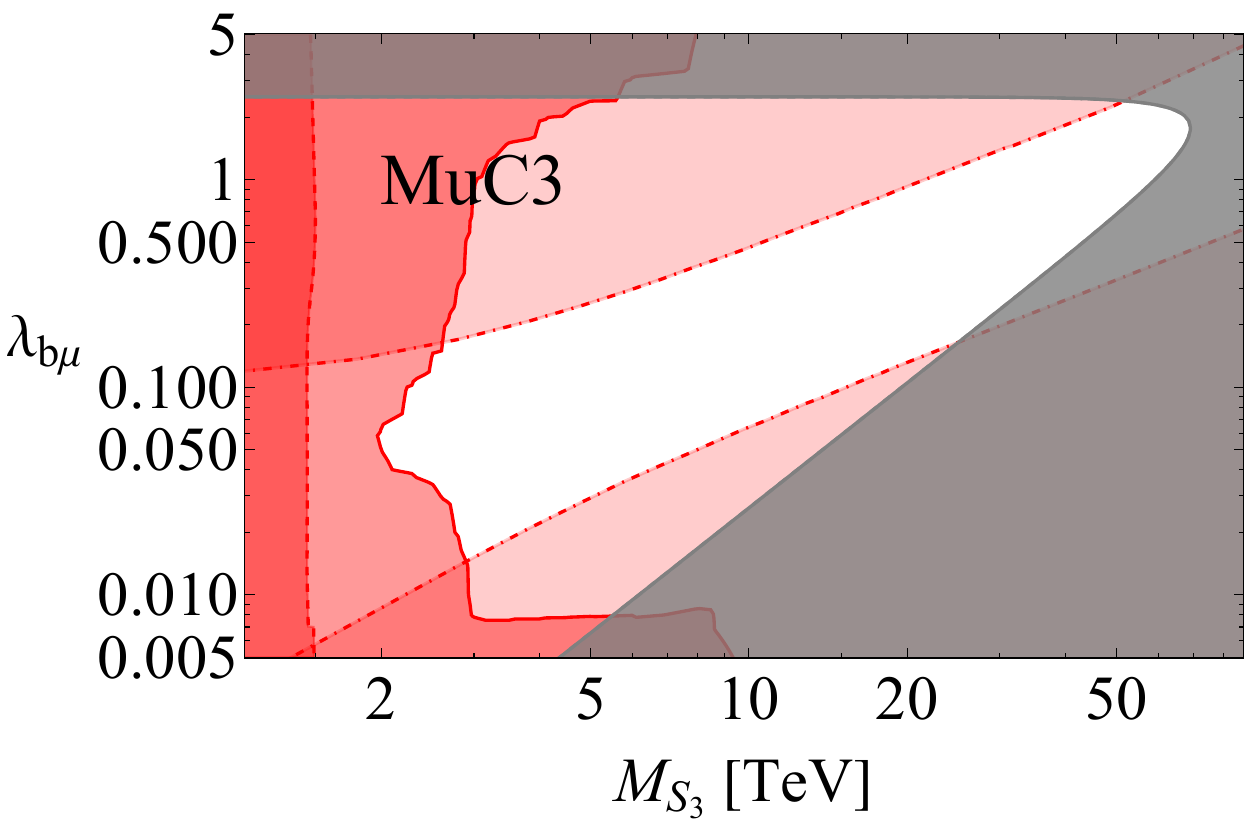}
\includegraphics[width=6.5cm]{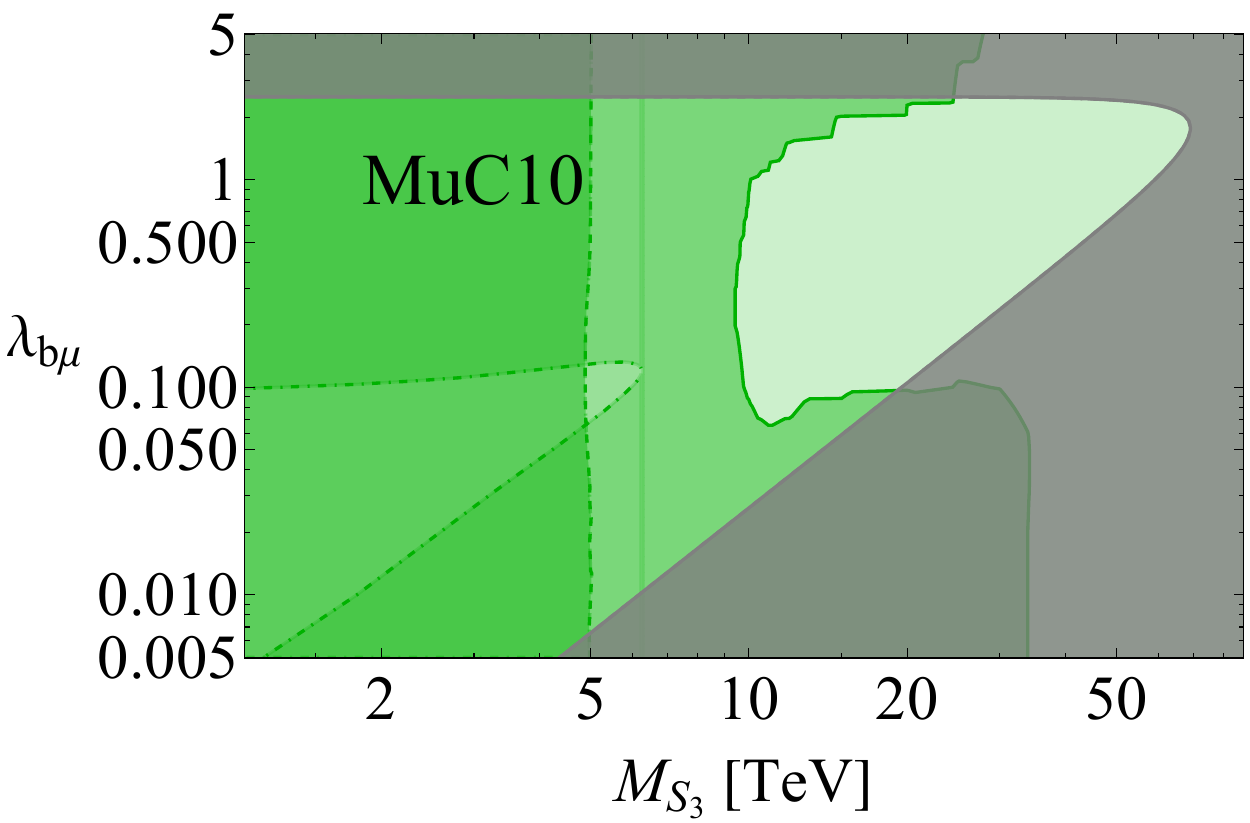}
\caption{\label{fig:S3_bounds} The 5$\sigma$ discovery prospects for the $S_3$ leptoquark behind to $bs\mu\mu$ anomalies, i.e. imposing Eq.~\eqref{eq:LQ_C9_num}. The present LHC exclusions at 95\%\,CL are shown as a thick black line. The dotted gray line corresponds to $\lambda_{b\mu} = \lambda_{s\mu}$. The perturbative limit $\Gamma_{S_3}/M_{S_3}<0.25$ is violated in the grey region. For future colliders, the discoverable region is on the side of the line where the corresponding label is. The smaller plots below the main one highlight the reach of various future colliders separately.}
\end{figure}

\subsubsection*{Addressing the $b s \mu \mu$ anomalies}

As shown by the extensive literature~\cite{Hiller:2014yaa,Buttazzo:2017ixm,Gripaios:2014tna,Gripaios:2015gra,Dorsner:2017ufx,Kumar:2018kmr,deMedeirosVarzielas:2019lgb,Crivellin:2019dwb,Saad:2020ihm,Gherardi:2020qhc,Lee:2021jdr,Marzocca:2021miv,Crivellin:2017zlb,Hiller:2017bzc,Angelescu:2021lln,Marzocca:2018wcf,Babu:2020hun,Saad:2020ucl,Greljo:2021xmg,Davighi:2020qqa}, $S_3$ is the only scalar leptoquark that can accommodate  $b s \mu \mu$ anomalies at the tree level. After integrating out $S_3$, we find the following contribution to the relevant effective operators
\begin{eqnarray}\label{eq:LQ_C9}
\Delta C_9^\mu = - \Delta C_{10}^{\mu} = \frac{ \pi}{\sqrt{2} G_F \alpha V_{ts}^* V_{tb}} \frac{\lambda_{b\mu} \lambda_{s\mu}}{M_{S_3}^2}~.
\end{eqnarray}
The fit to the $bs\mu\mu$ anomalies then implies
\begin{eqnarray}\label{eq:LQ_C9_num}
\lambda_{b\mu} \lambda_{s\mu} = 6.6 \times 10^{-4} \left( \frac{M_{S_3}}{\TeV} \right)^2 \left( \frac{\Delta C_9^\mu}{-0.39} \right)~.
\end{eqnarray}
In Fig.~\ref{fig:S3_bounds} we perform the collider sensitivity study in the $M_{S_3} - \lambda_{b\mu}$ plane, while fixing $\lambda_{s\mu}$ by Eq.~\eqref{eq:LQ_C9_num} where $\Delta C_9^\mu = - \Delta C_{10}^{\mu} = -0.39$. We do not show any complementary flavour physics constraints (such as the $B_s$ mixing) since those are loop suppressed in the leptoquark models and do not put limits on the parameter space of interest to this analysis (for a global fit with the $S_3$ see Ref.~\cite{Gherardi:2020qhc} and with the $U_1$ see Ref.~\cite{Cornella:2021sby}).

The present LHC bounds at 95\%\,CL from the DY process and the leptoquark pair production are shown with thick black lines. The $5\sigma$ discovery prospects for future colliders are depicted with various colored lines. The corresponding label for a collider and a process is always on the excluded (or discoverable) side.
Again, we report four small dedicated sensitivity plots for each collider for an easier comparison.
The HL-LHC can not discover much more of the parameter space that is not already excluded. However, the FCC-hh will explore all but a fraction of the parameter space in between the dashed purple (DY) lines and the vertical solid purple line (pair-production). This region of parameter space  corresponds to $\lambda_{b\mu} \approx \lambda_{s\mu}$, which minimizes the contribution to $pp \to \mu\mu$ once Eq.~\eqref{eq:LQ_C9_num} is imposed and is also beyond the pair production reach for higher masses. Note that this region of parameter space strongly violates the $U(2)^3$ flavour symmetry in the quark sector. In this case, if the new physics sector is rich with resonances, then there remains a puzzle to be explained:  the absence of flavour changing neutral currents. 

Moving on to muon colliders, a 3 TeV MuC discovery reach in the IDY channel is comparable to the one of the FCC-hh. However, the leptoquark pair production prospects are substantially lower, stopping at $M_{S_3} \approx \sqrt{s_0}/2 = 1.5~\rm TeV$ which is even below the present LHC exclusion. On the other hand, the MuC10 will test the whole parameter space by combining different channels: IDY, pair production, and $\mu q \to \mu j$. Interestingly, both a 3\,TeV and a 10\,TeV MuC might directly observe an $s$-channel resonance in the $\mu q \to \mu j$ (see Section~\ref{sec:resonant_production}) for masses up to approximately $\sqrt{s_0}$. In other words, this seems to be the most promising on-shell process at muon colliders.

\subsection{Vector leptoquark $U_1$}
\label{sec:LQvector}

\begin{figure}[t]
\centering
\includegraphics[width=13cm]{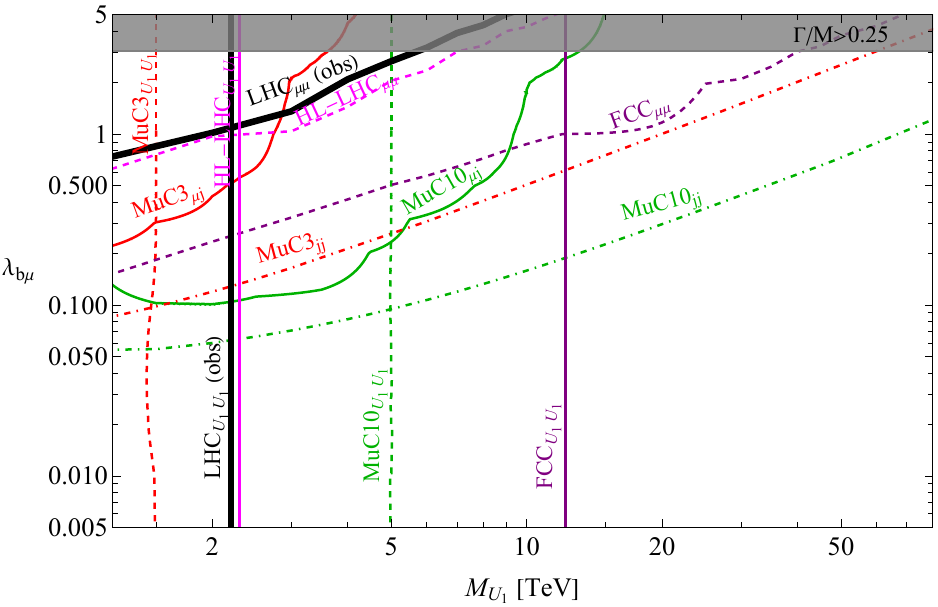}
\caption{\label{fig:U1_bounds_bmu} Discovery reach at 5$\sigma$ for the $U_1$ leptoquark in the $U(2)^3$ symmetric case. The present 95\%CL exclusion by LHC is shown as a thick black line. In the grey region, the perturbativity limit $\Gamma_{U_1}/M_{U_1}<0.25$ is violated.}
\end{figure}

\begin{figure}[t]
\centering
\includegraphics[width=13cm]{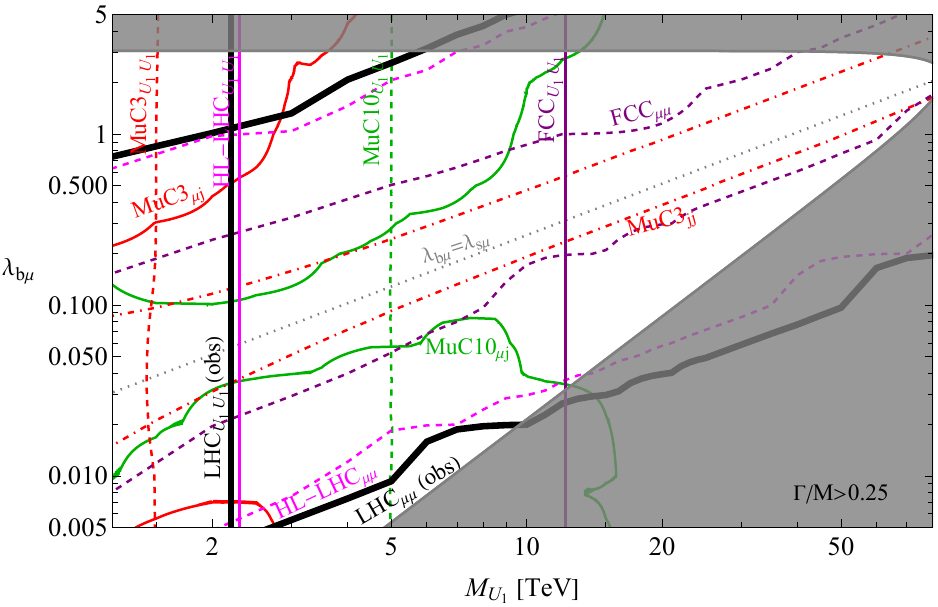} \\
\includegraphics[width=6.5cm]{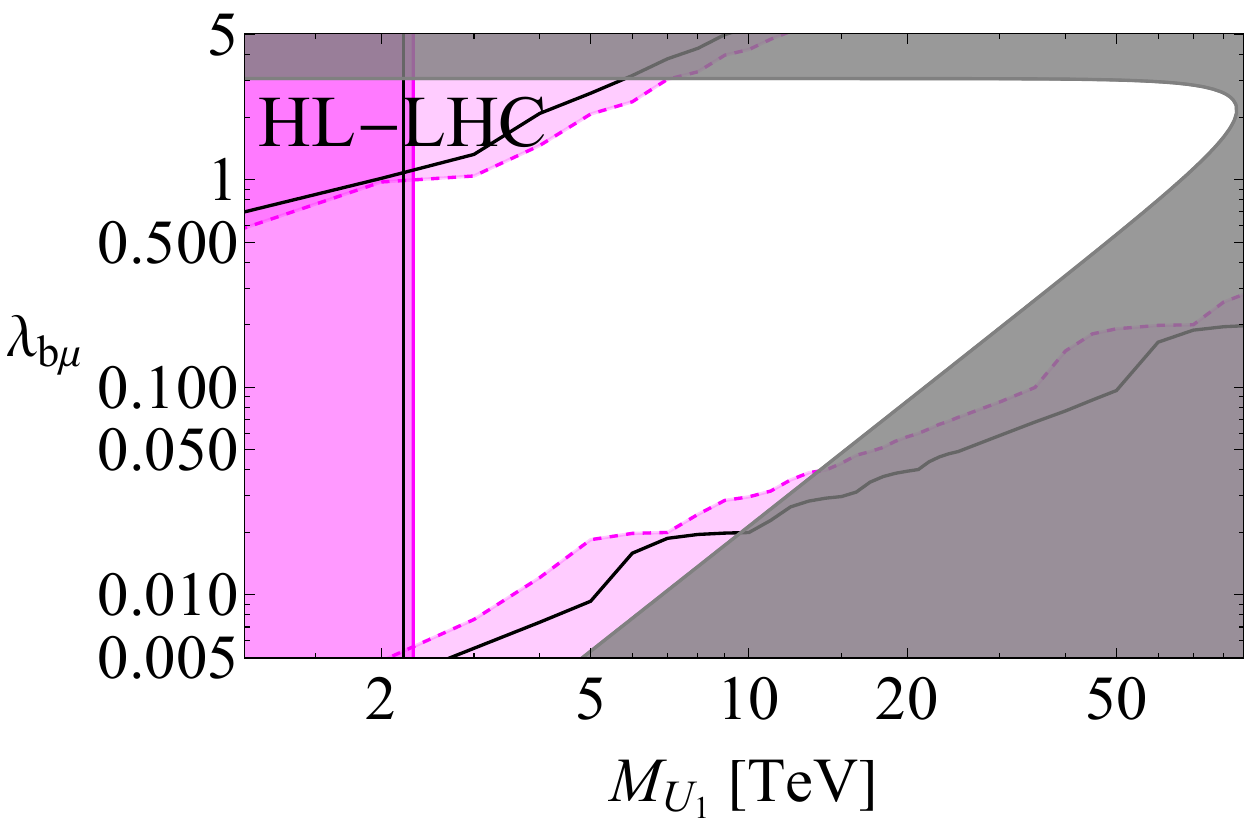}
\includegraphics[width=6.5cm]{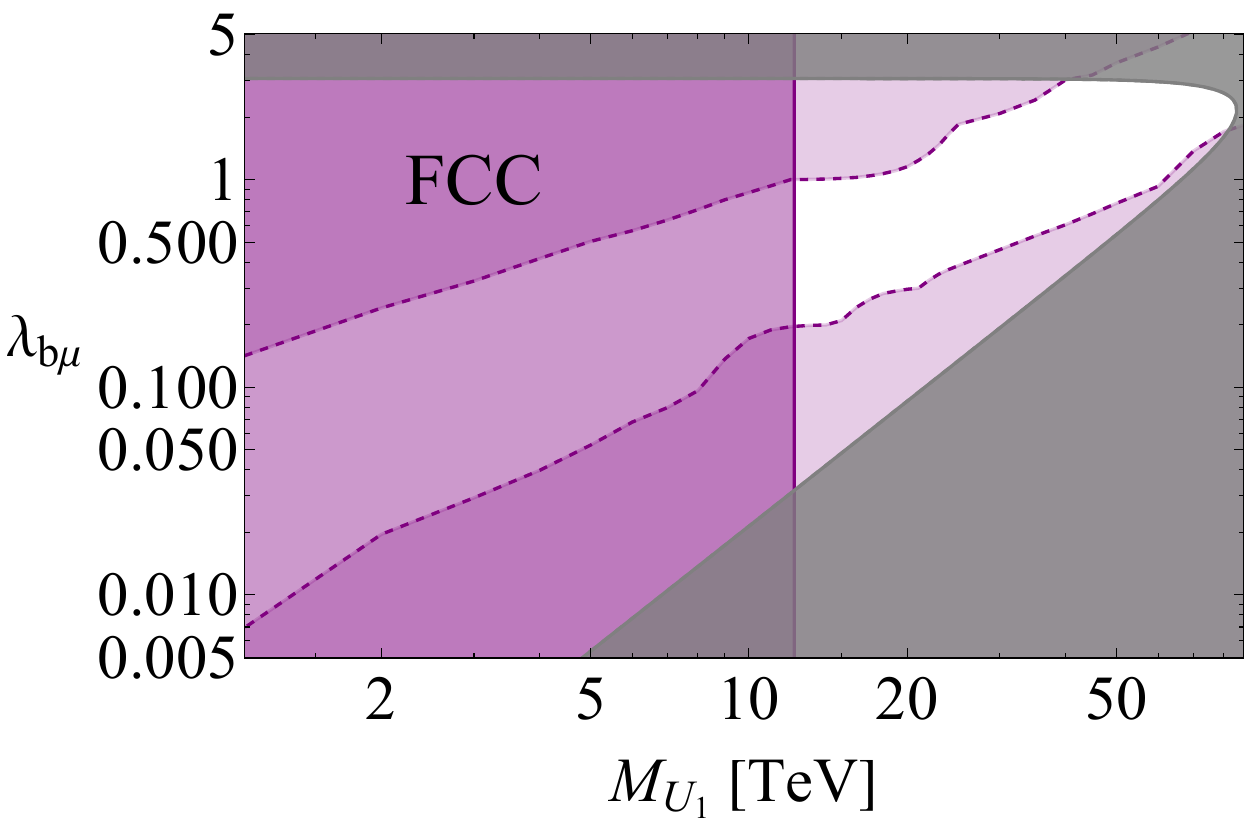} \\
\includegraphics[width=6.5cm]{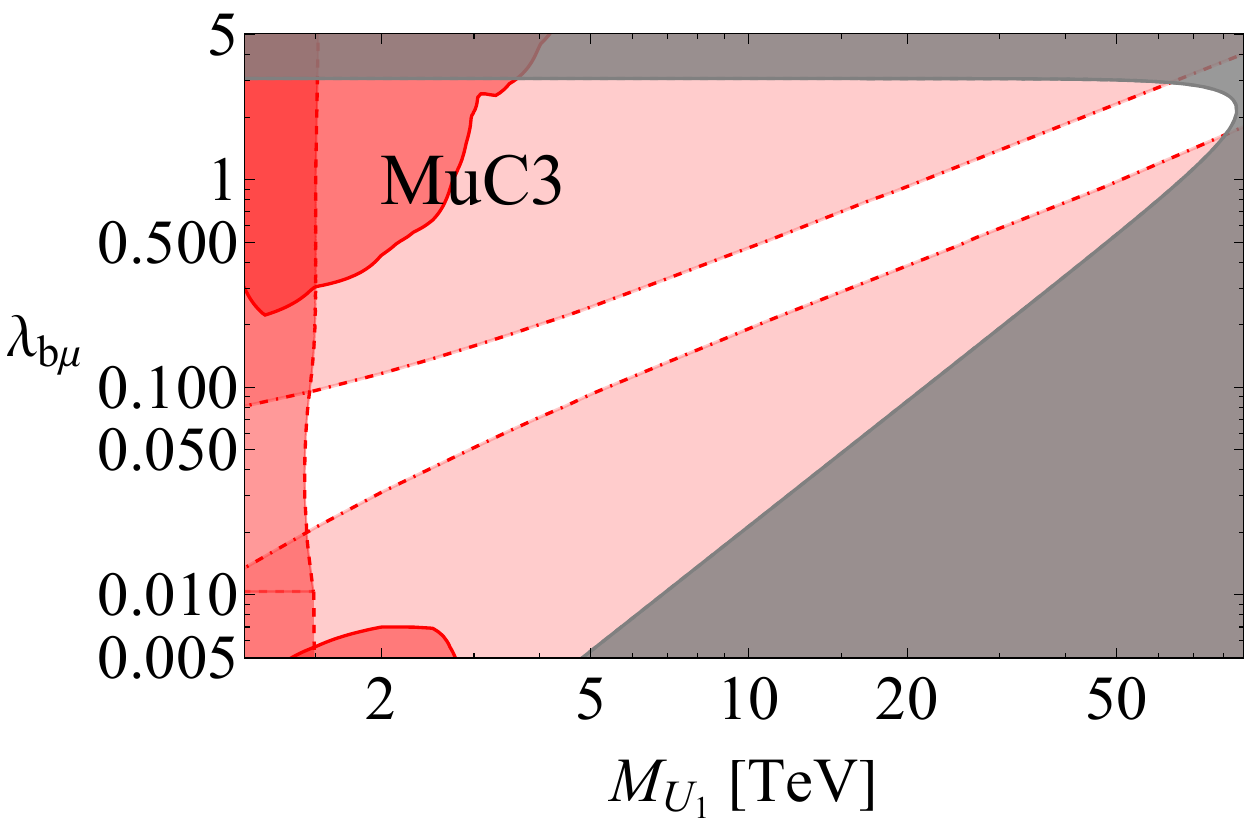}
\includegraphics[width=6.5cm]{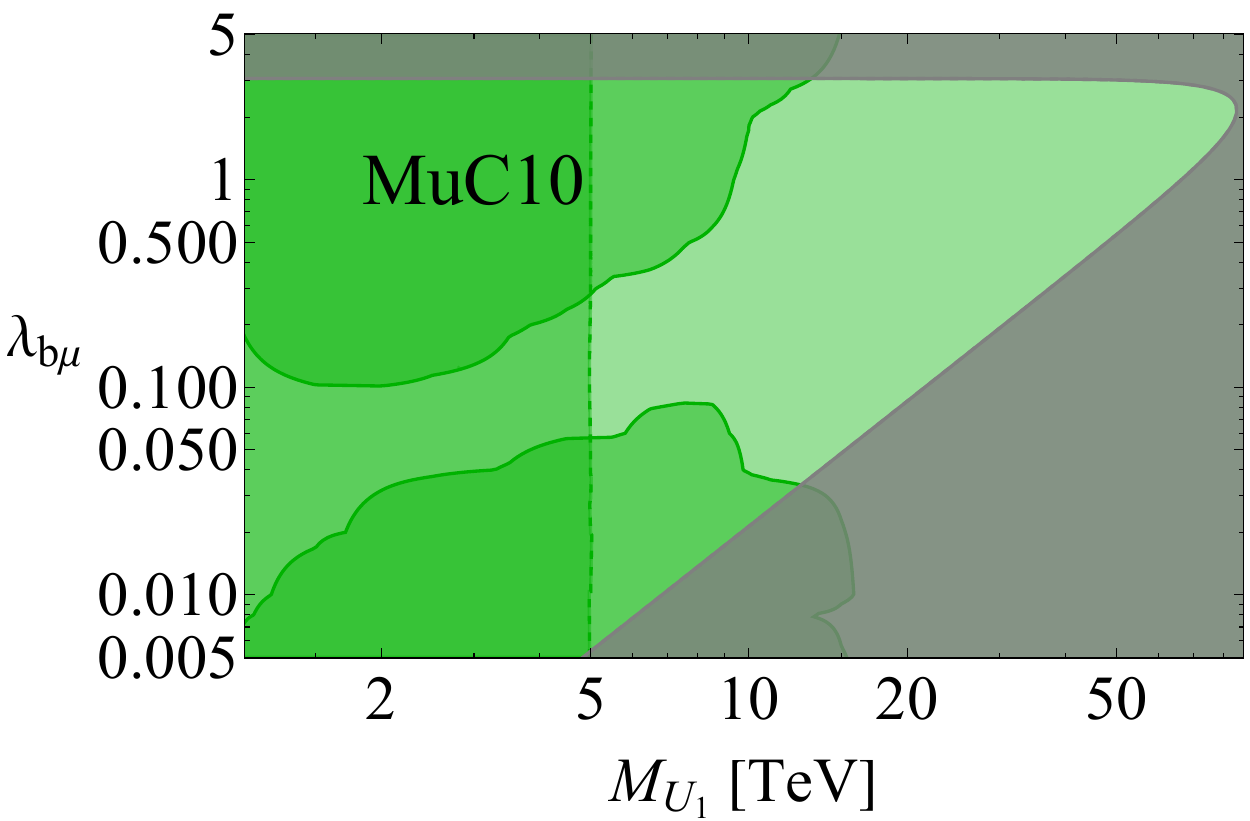}
\caption{\label{fig:U1_bounds} Discovery reach at 5$\sigma$ for the $U_1$ leptoquark. The fit to $bs\mu\mu$ anomalies is imposed everywhere via Eq. \eqref{eq:LQ_C9_num}. The present 95\%CL exclusion by LHC is shown as a thick black line. The dotted gray line corresponds to $\lambda_{b\mu} = \lambda_{s\mu}$.
In the grey region, the perturbativity limit $\Gamma_{U_1}/M_{U_1}<0.25$ is violated. The limit for the  {\boldmath$\mu \bar{\mu}$}$\to j j$ channel at 10 TeV MuC covers the whole plane. For future colliders, the discoverable region is the one on the side of the line where the corresponding label has been drawn. The smaller figures below the main figure highlight a single future collider at a time.}
\end{figure}

Let us consider extending the SM with a heavy vector leptoquark $U_1\sim ({\bf 3},{\bf 1},2/3)$~\cite{Dorsner:2016wpm}.
Assuming only left-handed couplings, the interaction Lagrangian is
\begin{eqnarray}
{\cal L}_{U_1}^{\rm int} = \lambda_{i\mu}  \overline{Q_L}^i \gamma_\alpha L_L^2 U_1^\alpha + \text{h.c.} = \lambda_{i\mu} U_1^\alpha \left( V_{ji} \bar{u}_L^j \gamma_\alpha \nu_\mu + \bar{d}_L^i \gamma_\alpha \mu_L \right) + \text{h.c.}~,
\end{eqnarray}
while interactions with the SM gauge bosons are described by the Lagrangian
\be
    \mathcal{L}_{U_1}^{\rm gauge} = - \frac{1}{2} U_{\mu\nu}^\dagger U^{\mu\nu} - i g_s \kappa_s U_{1 \mu}^\dagger T^a U_{1 \nu} G^{a \mu\nu} - i g^\prime \frac{2}{3} \kappa_Y U_{1 \mu}^\dagger U_{1 \nu} B^{\mu\nu} ~,
    \label{eq:U1gaugekappa}
\ee
where $U_{\mu\nu} = D_{\mu} U_{1\nu} - D_{\nu} U_{1\mu}$. The dimensionless parameters $\kappa_{s,Y}$ depend on the specific UV completion of the model. We assume that $U_{1\mu}$ arises from a spontaneously broken Yang-Mills theory, that is $\kappa_{s,Y} = 1$.\footnote{We note that the for the benchmark $\kappa_{Y} = 0$ our phenomenological analysis remains identical, since the main production and decay channels are dominated by the QCD and $q\ell$-LQ couplings. Instead, $\kappa_{s} = 0$ reduces the pair production cross section, reducing the $U_1 U_1$ limits from hadron colliders by about $30\%$ on the mass, see e.g.~\cite{Dorsner:2018ynv}.} The $U_1$ decay width is
\be
    \Gamma_{U_1} = \frac{|\lambda_{b\mu}|^2 + |\lambda_{s\mu}|^2}{12\pi} M_{U_1}~,
\ee
where we take only $\lambda_{b\mu}$ ($i=3$) and $\lambda_{s\mu}$ ($i=2$) couplings to be non-zero. Analogously to the $S_3$ case, we consider two different scenarios for the $\lambda_{s\mu}$ coupling.

\subsubsection*{$U(2)^3$ symmetric case}

The minimally broken $U(2)^3$ quark flavour symmetry predicts $|\lambda_{s\mu}| \sim |V_{ts} \lambda_{b\mu}|$ and even smaller $|\lambda_{d\mu}| \sim |V_{td} \lambda_{b\mu}|$, making all leptoquark couplings but $\lambda_{b\mu}$ irrelevant for high-$p_T$ processes. On the leptonic side, we again assume an exclusive coupling to muons.

The present LHC constraints, as well as future $5\sigma$ discovery prospects are presented in the $M_{U_1} - \lambda_{b\mu}$ plane in Fig.~\ref{fig:U1_bounds_bmu}. The resulting picture is qualitatively similar to the one for the $S_3$ leptoquark in Section~\ref{sec:LQscalar}, to which we refer for a detailed discussion. The extra interaction term with gluons increases the reach at FCC-hh from the pair production up to $M_{U_1} \approx 12.9 \,\TeV$, while the reach at MuC is limited to $\sqrt{s_0}/2$ by kinematics. Note also the suppression in reach from the $\mu j$ final state at MuC, compared to the scalar case. This is due to a factor of $1/2$ from the branching ratio of $U_1 \to d^i \mu$ and a factor of $\sqrt{2}$ less in the coupling to $d^i \mu$.
As before, the MuC10 sensitivity prospect in the IDY channel is almost an order of magnitude better than at FCC-hh, while the MuC3 is comparable.

\subsubsection*{Addressing $bs\mu\mu$ anomalies}

The $U_1$ vector leptoquark has attracted a great deal of attention in the context of the $B$-meson anomalies~\cite{Barbieri:2015yvd,Alonso:2015sja,Calibbi:2015kma,Bhattacharya:2016mcc,Buttazzo:2017ixm,DiLuzio:2017vat,Greljo:2018tuh,Bordone:2017bld,Bordone:2018nbg,Cornella:2019hct,Fornal:2018dqn,Blanke:2018sro,Fuentes-Martin:2019ign,Guadagnoli:2020tlx,Fuentes-Martin:2020bnh,Angelescu:2021lln,Fuentes-Martin:2019bue,Bernigaud:2021fwn,Fuentes-Martin:2020luw,Fuentes-Martin:2020hvc,Fuentes-Martin:2022xnb,DiLuzio:2018zxy}. Regarding the contribution to $\Delta C_9^\mu = - \Delta C_{10}^\mu$, the expressions in Eqs. \eqref{eq:LQ_C9} and \eqref{eq:LQ_C9_num} hold identically with a replacement $M_{S_3} \to M_{U_1}$. In the following, we focus on the parameter space that can address the $b s \mu \mu $ anomalies. In particular, $\lambda_{s\mu}$ is fixed by the best-fit point from the $bs\mu\mu$ analysis, see Eq.~\eqref{eq:LQ_C9_num}, leaving only two input parameters $M_{U_1}$ and $\lambda_{b\mu}$.

The results of our collider studies are summarised in Fig.~\ref{fig:U1_bounds}, where the color coding is the same as before. The prospects at the hadron colliders are similar to the $S_3$ case, see Section~\ref{sec:LQscalar} for a detailed discussion. At MuCs, the IDY channel provides somewhat higher sensitivity, unlike the resonant production in $\mu q \to \mu j$. The MuC10 can discover at $5\sigma$ the entire viable parameter space by the IDY process alone, while the MuC3 can exclude it at 95\% CL. However, it is important to stress that this is only an indirect effect: the leptoquark is exchanged in the $t$-channel and therefore a smooth distortion is predicted (see Fig.~\ref{fig:S3XsecMuC}). Should the effect in the IDY be discovered, the characterisation of the new physics will become a challenge. This is to be compared with FCC-hh, where the vector leptoquark can be discovered as an on-shell resonance in the pair production process up to $M_{U_1} \approx 12.9 \,\TeV$.

\section{Conclusions}
\label{sec:conc}

\begin{figure}[t]
\centering
\includegraphics[width=13cm]{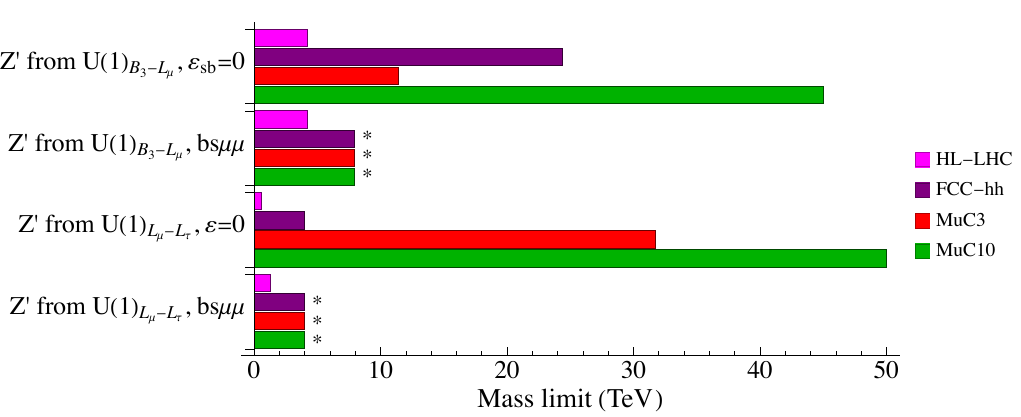} \\
\includegraphics[width=13cm]{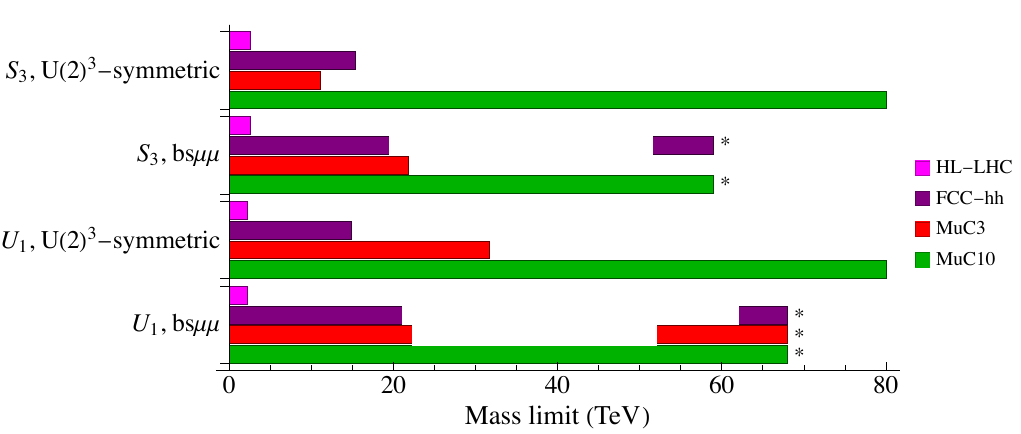}
\caption{\label{fig:model_comparison} Comparison of the discovery reach of future colliders for the different models discussed in this work. The corresponding couplings ($g_{Z^\prime}$ or $\lambda_{b\mu}$) are set to $1$. We denote with an asterisk cases where the reach exceeds the perturbativity limit or meet any other relevant bound (e.g. from meson mixing).}
\end{figure}

The near-term future of particle physics will be charted by precision measurements. The LHC has (almost) reached its nominal beam energy and is marching towards the high-luminosity phase. Within this decade, the Belle II experiment in Japan plans to surpass the previous $B$ factories by one order of magnitude in integrated luminosity. It could very well be that a new physics will first show up indirectly in the precision measurements at low energies. Interesting (though yet indecisive) hints are reported by the LHCb collaboration, suggesting a microscopic new physics in $b \to s \mu^+ \mu^-$ decays.

The long-term future of the field crucially depends on the decisions we make today about the next generation of high-energy colliders. The two most prominent options on the table are the future proton collider at 100~TeV (FCC-hh) and a multi-TeV muon collider (MuC). It is, therefore, necessary to thoroughly compare the two on a broad set of new physics hypotheses. In this work, we consider a short-distance new physics relevant for rare (semi)leptonic $B$ meson decays and investigate the complementary discovery prospects at future colliders specified in Table~\ref{tab:colliders}. Our set of benchmarks includes heavy $Z'$ and leptoquark mediators as well as semileptonic four-fermion interactions.

Our MuC studies focus on two-body final states sensitive to such new physics, the most relevant being di-jet, di-muon, di-tau, and muon plus jet (see Fig.~\ref{fig:FDiagrams}).
For each topology we compute the leading SM background as well as the expected new physics signal (see Figs.~\ref{fig:ZprimeXsecMuC} and \ref{fig:S3XsecMuC} for examples of differential cross sections).
We identify an intriguing interplay between resonant and non-resonant signal shapes relevant to the discovery prospects (see Fig.~\ref{fig:ZprimeB3Lmu_mumujj}). One of the novelties of our work is the employment of precise electroweak parton distribution functions for the high-energy muon beam, as explained in Appendix~\ref{app:MuonPDFs}. This has several interesting phenomenological implications, defying common intuition from the LHC.  The complementary processes considered at hadron colliders are summarised in Fig.~\ref{fig:FCCDiagrams}. 
These include di-muon and four-muon final states.
Furthermore, for both muon and hadron colliders we also study the sensitivity from leptoquark pair production.

In our first analysis we assume that the new mediators are too heavy for on-shell production even at future colliders and study the deviations in the high-energy tails due to new semileptonic four-fermion interactions (Section~\ref{sec:contact}). For example, in Fig.~\ref{fig:EFTlimits_SU3} we compare the projections from the di-jet final state at muon colliders to the di-muon and muon plus neutrino final states at hadron colliders, on operators involving left-handed doublets with different $SU(2)_L$ contractions, $(\bar L^2_L \gamma_\alpha L^2_L) ( \bar Q^i_L \gamma^\alpha Q^j_L )$ and $(\bar L^2_L \gamma_\alpha \sigma^a L^2_L) ( \bar Q^i_L \gamma^\alpha \sigma^a Q^j_L )$, assuming flavour universality in the quark-sector.
Interestingly, in this case a 3\,TeV MuC only slightly improves upon the HL-LHC, while the FCC-hh probes substantially more parameter space, that can only be matched by a 14\,TeV MuC.
We then study similar operators but with a different flavour structure: $ b s \mu \mu $ and $b b \mu \mu$. The corresponding results are shown in Figs.~\ref{fig:EFTlimits_bs} and \ref{fig:EFTlimits_bb}.
In this case, even a 3\,TeV MuC provides comparable sensitivity to the FCC-hh, while 10\,TeV and 14\,TeV MuCs are considerably better. This is due to the suppression from the sea quark parton distribution functions at hadron colliders.
In the same figures we also indicate a tentative prediction from the present-day LHCb anomalies in $b \to s \mu^+ \mu^-$ decays. Both MuC3 and FCC-hh have a good prospect to reach this target, assuming realistic models in which the $b b \mu \mu$ interactions are larger than the $b s \mu \mu$. Instead, the MuC10 (MuC14) can easily cover even the most pessimistic case compatible with the LHCb anomalies.

A heavy $Z^\prime$ vector arises in several new physics models. In our work we consider a $Z'$ associated with the gauging of the $U(1)_{B_3 - L_\mu}$ (Section~\ref{sec:ZprimeB3Lmu}) or the $U(1)_{L_\mu - L_\tau}$ (Section~\ref{sec:ZprimeLmuLtau}) group.
While a coupling to muons is present by design in both cases, the $Z'$ interaction with quarks is very different. For both models we consider two scenarios: $i)$ the set of renormalisable couplings predicted by the corresponding $U(1)_X$ and $ii)$ a minimal extension needed to address the $b s \mu \mu$ anomalies. Our main findings are presented in Figs.~\ref{fig:Zp_B3-Lmu_bounds_noRK} and \ref{fig:Zp_B3-Lmu_bounds} for $B_3 - L_\mu$ and  Figs.~\ref{fig:Zp_Lmu-Ltau_bounds_noRK} and \ref{fig:Zp_Lmu-Ltau_bounds} for $L_\mu - L_\tau$. In the scenario $i)$, the FCC-hh compares to the MuC10 for the first model, while in the second model (which is quark-phobic), even a 3\,TeV MuC is better than the FCC-hh. When the $ b s \mu \mu $ anomalies are addressed, we find that the whole viable parameter space in both models (with resonances above the EW scale) can be fully explored at the MuC3 as well as the FCC-hh. In case of $L_\mu - L_\tau$ at hadron colliders, this is possible thanks to the $p p \to \mu^+ \mu^- Z' \to 4 \mu$ process. Our results are also summarized in Fig. \ref{fig:model_comparison}, where one can directly compare the $Z^\prime$ mass reach of the future colliders for a  benchmark value of the coupling $g_{Z^\prime} = 1$.

The interest in leptoquark models at the TeV scale and their phenomenology has received a boost since the appearance of the flavour anomalies. Leptoquarks are also motivated by the hinted quark-lepton unification. In this paper, we study both scalar (Section~\ref{sec:LQscalar}) and vector (Section~\ref{sec:LQvector}) leptoquarks, with couplings only to the second generation lepton doublet.
In both cases, we consider two possible flavour structures for the couplings to quarks: $i)$ one dictated by an exact $U(2)_Q$ flavour symmetry and $ii)$ a minimal scenario required to address the $b s \mu \mu$ anomalies.
The main results are given in Figs.~\ref{fig:S3_bounds_bmu} and \ref{fig:S3_bounds} for the scalar, and Figs.~\ref{fig:U1_bounds_bmu} and \ref{fig:U1_bounds} for the vector, with similar conclusions. Moreover, Fig.~\ref{fig:model_comparison} offers again a simplified overview of the results for a choice of coupling $\lambda_{b\mu} = 1$.
Both MuC3 and FCC-hh have the potential to cover large portions of the parameter space, with similar sensitivity on the leptoquark coupling while the FCC-hh has a much better prospect for on-shell leptoquark discovery via pair-production. The sensitivity at the MuC3 indeed comes almost completely from off-shell effects in the di-jet final state.
This aspect is improved at the MuC10 (and even more so at the MuC14). Indeed, among the colliders considered here, a 10 TeV (or higher) MuC is the only one that can cover (with a 5$\sigma$ discovery prospect) the whole parameter space viable for addressing the $bs\mu\mu$ anomalies with leptoquarks.

\section*{Acknowledgments}

The work of AG and JS has received funding from the Swiss National Science Foundation (SNF) through the Eccellenza Professorial Fellowship ``Flavor Physics at the High Energy Frontier'' project number 186866. AA, DM, and ST acknowledge support by MIUR grant PRIN 2017L5W2PT. The work of AG and DM is also partially supported by the European Research Council (ERC) under the European Union’s Horizon 2020 research and innovation programme, grant agreement 833280 (FLAY).

\appendix

\section{Muon PDFs}
\label{app:MuonPDFs}

Initial state radiation in high-energy lepton colliders has an important impact on the phenomenology of such a machine: it spreads the lepton energy to lower values and generates different possible initial states. In QED, a lepton can emit a photon, that in turn can split into $\ell^+ \ell^-$ pair. The probability of these processes increases logarithmically with the energy~\cite{vonWeizsacker:1934nji,Williams:1934ad}. Thus, for large enough energies they must be resummed following the analogous of the Dokshitzer-Gribov-Lipatov-Altarelli-Parisi (DGLAP) equations for QCD~\cite{Gribov:1972ri,Dokshitzer:1977sg,Altarelli:1977zs}. The resulting PDFs describe the probability of finding a certain parton `$i$', with a given fraction $x$ of the initial momentum, inside the original lepton in a process of energy $Q$: $f_i(x,Q)$. For energies much above the EW scale, EW PDFs with the complete set of SM interactions must be taken into account \cite{Ciafaloni:2005fm,Chen:2016wkt,Bauer:2017isx,Bauer:2018arx}.

\begin{figure}
	\centering
	\includegraphics[scale=1, width= 0.7\textwidth]{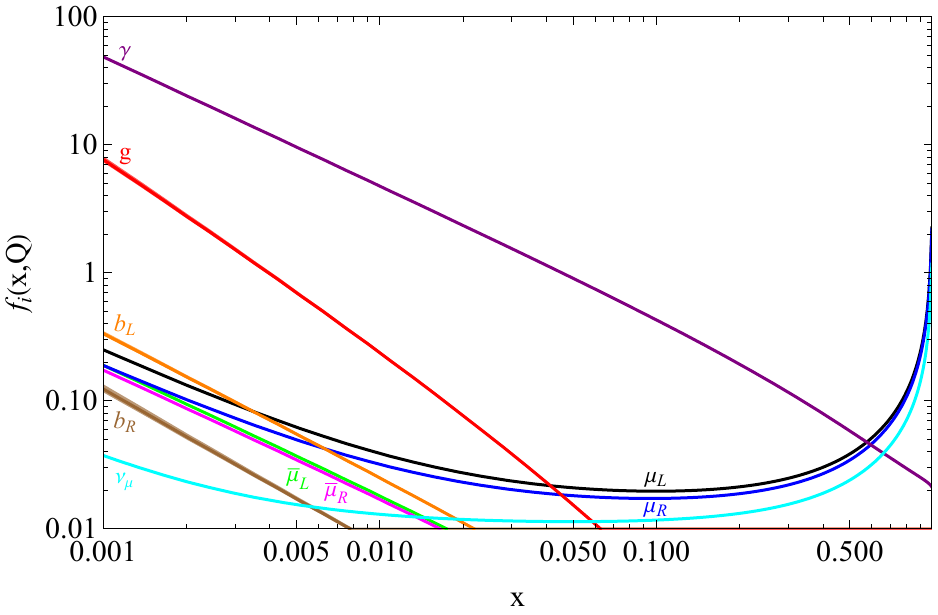}
	\caption{Muon PDFs including the full unbroken SM interactions for $Q = 3~\TeV$. The thickness of the gluon and $b$ quarks PDFs is obtained by varying the $\mu_{QCD}$ scale from 0.5 to 1 GeV.}\label{fig:pdfsmub}
\end{figure}

In practice, for our MuC analysis we derive muon PDFs by resumming soft real emissions as well as virtual radiation, needed to cancel soft divergences, by numerically solving the DGLAP equations for the system with an initial condition $f_\mu(x, m_\mu) = \delta(1-x)$ and $f_{i \neq \mu}(x, m_\mu) = 0$.
We follow the strategy laid out in \cite{Han:2020uid,Han:2021kes}. In the first phase of the evolution, below the electroweak scale $Q = \mu_{EW}\sim M_Z$, we consider only QED and QCD interactions involving the three leptons and five quark flavours (top quark excluded), with QCD contribution starting at the scale $Q = \mu_{QCD}\sim 0.7\, \textrm{GeV}$ (see Ref.~\cite{Drees:1994eu}).\footnote{We evaluate the uncertainty due to the choice of this scale by changing $\mu_{QCD}$ from 0.5 to 1 GeV. As a result, we find a $\approx 10\%$ variation in the quark and gluon PDFs at the TeV scale and for $10^{-3} \lesssim x < 1$, while the impact on leptons or EW gauge bosons is completely negligible ($\sim \mathcal{O}(10^{-5})$). This is shown in Fig.~\ref{fig:pdfsmub}.} In this phase, thanks to the vectorlike nature of the theory, we do not separate fermion chiralities or vector polarizations.
Then at the electroweak scale $\mu_{EW}$ we match with the PDFs obtained from the first phase and continue the evolution considering the full unbroken Standard Model interactions, now separating left and right chiralities, as well as the two transverse polarizations of gauge bosons. The role of the longitudinal polarization is played by the Goldstone bosons coming from the Higgs doublet and we identified PDFs with the same equations and initial conditions.
Since for this work we need only muon, neutrino, and quark PDFs, we neglect ultra-collinear effects arising in the broken phase. These are suppressed as $m_W^2 / Q^2$ at higher energies but give the dominant contribution to longitudinal gauge bosons PDFs. However, their impact on massless fermions is negligible. We leave the implementation of these effects to an upcoming work on muon PDFs, where all other details of this computation will be available \cite{muPDF}.

\begin{figure}
	\centering
	\includegraphics[scale=1, width= 0.48\textwidth]{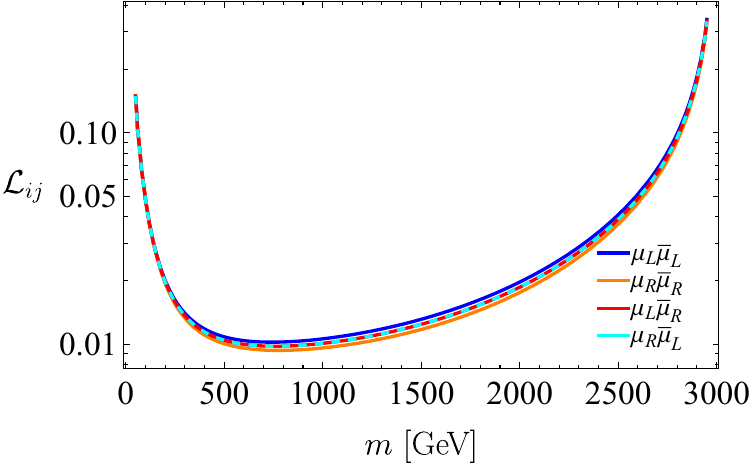}
	\includegraphics[scale=1, width= 0.48\textwidth]{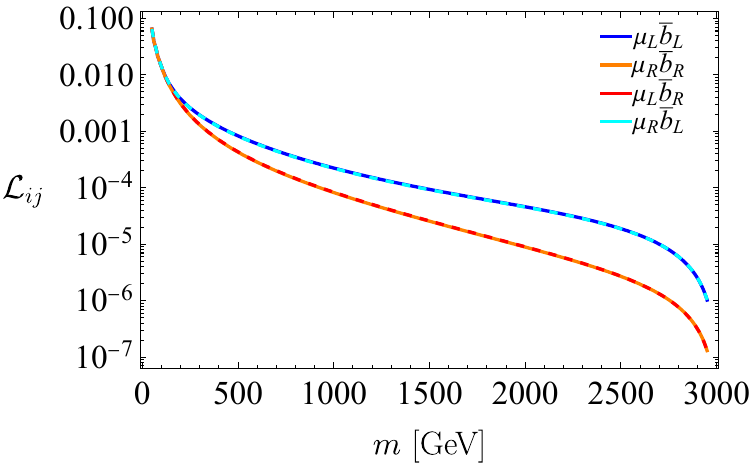}
	\caption{Parton luminosities, for $\sqrt{s_0}=3~\TeV$, involving two muons (left) or a muon and a $\bar{b}$ (right).}\label{fig:lumimumu}
\end{figure}

In Fig.~\ref{fig:pdfsmub} we report the PDFs relevant for our work at the scale $Q = 3\, \textrm{TeV}$, while in Fig.~\ref{fig:lumimumu} we show the parton luminosities $\mathcal{L}_{ij}$ used to compute the various cross sections in this work, given by
\begin{equation}\label{eqA:lumin}
\mathcal{L}_{ij}(\tau) = \int_\tau^1 \frac{dx}{x} f_i(x, m) f_j\left(\frac{\tau}{x}, m\right),
\end{equation}
where $f_i(x, m)$ is the PDF of the parton $i$ computed at a scale $Q = m$ and $\tau$ is defined as
\begin{equation}
\tau = \frac{m^2}{s_0},
\end{equation}
with $m$ being the invariant mass of the two initial states and $s_0$ the center of mass energy of the collider.
By comparing  our results with Fig.~1 of \cite{Han:2021kes} we find good agreement for all PDFs considered in this work (fermions, gluon and photon), with deviations of $\lesssim 10\%$.

The luminosity is related to a probability for a collision between partons $i$ and $j$ with energy $\sqrt{\tau s_0}$. For a given process, the total cross section is obtained after a convolution with a partonic cross section,
\begin{equation}
\sigma_{TOT} = \sum_{i,j}\int_0^1d\tau\mathcal{L}_{ij}(\tau)\sigma_{ij}(\sqrt{\tau s_0}) = \sum_{i,j}\int_0^{\sqrt{s_0}} dm\frac{2m}{s_0}\mathcal{L}_{ij}\left(\frac{m^2}{s_0}\right)\sigma_{ij}(m).
\end{equation}

\section{Partonic cross sections}
\label{app:partonic_xsec}

In a scattering process where partons of type 1 collide with partons of type 2 to produce partons of type 3 and 4, the differential cross section defined in the lab frame is given by
\begin{equation}
    \frac{d^3\sigma}{dy_3 dy_4 dm}= f(x_1) f(x_2) \frac{m^3}{2s}\frac{1}{\cosh y_*}\frac{d\sigma}{d \hat t} (1+2 \to 3+4)~,
    \label{eq:triplediffXsec}
\end{equation}
where $m$ is the invariant mass of the products, $y_i$ is the rapidity of parton $i$, $f(x_i)$ is the PDF and
\begin{align}
    x_{1,2}=\frac{m}{\sqrt{s_0}}e^{\pm \frac{y_3+y_4}{2}}~, \quad y_*=\frac{1}{2}(y_3-y_4)~, \quad
    \hat t = -\frac{m^2}{2} (1-\cos\theta_*)~, \quad
    \theta_*=\arcsin\left(\frac{1}{\cosh y_*}\right)~.
\end{align}
For the process $\mu_L^+\mu_L^- \to \bar q_L q_L$ we derive analytic expressions for the total polarized partonic cross sections (not averaged over initial spins). As an example, we give here the result in the limit of vanishing fermion masses for the case where the NP effect is mediated by a $Z^\prime$, a $S_3$ leptoquark, as well as a contact interaction $C_{\mu_L \mu_L q_L q_L}$ in Eq.~\eqref{eq:EFT_collider}:
\begin{align}
&\sigma(\mu_L^+\mu_L^- \to \bar q_L q_L) = \frac{3}{16 \pi s^2}\bigg[ \frac{4 s^3}{3} \left| \frac{Q_q Q_\mu}{s} + \frac{ g^Z_q g^Z_\mu}{s-m_Z^2 + i \Gamma_{Z} m_{Z} } + \frac{g^{Z^\prime}_{q_L} g^{Z^\prime}_{\mu_L}}{s-M_{Z^\prime}^2 + i \Gamma_{Z^\prime} M_{Z^\prime}} + C_{\mu_L \mu_L q_L q_L} \right|^2 \notag \\ 
&-2 \left|\lambda_{q\mu}\right|^2 \Re \left[\left( \frac{Q_q Q_\mu}{s} + \frac{g^Z_q g^Z_\mu}{s-m_Z^2+ i \Gamma_{Z} m_{Z}}\right) \left( 2 M_{S_3}^2 s - s^2 +2 M_{S_3}^4 \log  \left(\frac{M_{S_3}^2}{M_{S_3}^2+s}\right) \right) \right] \notag \\
& \left. +\left|\lambda_{q\mu}\right|^4 \left(\frac{s(2 M_{S_3}^2+s)}{M_{S_3}^2+s} +2 M_{S_3}^2 \log \left(\frac{M_{S_3}^2}{M_{S_3}^2+s}\right) \right) \right]~,
\end{align}
where the equation above can be used directly for the up quarks with the coupling convention of Eq.~\ref{eq:S3LQcoup}, while for the down quarks one should multiply the leptoquark coupling by a factor of $\sqrt{2}$ (see Eq.~\eqref{eq:S3_def}).
The cross section for the vector leptoquark is:
\begin{align}
&\sigma(\mu_L^+ \mu_L^-\to q_L q_L^-)=\frac{3}{16\pi s^2}\left[ \frac{4 s^3}{3} \left|\frac{Q_q Q_\mu}{s}+\frac{g_q^Z g_q^\mu}{s-m_Z^2+ i \Gamma_{Z} m_{Z} }\right|^2\right.
\notag\\
&+ 4|\lambda_{q\mu}|^2 \Re\left[\left(\frac{Q_q Q_\mu}{s}+\frac{g_q^Z g_q^\mu}{s-m_Z^2+ i \Gamma_{Z} m_{Z}}\right) \left( 2 M_{U_1}^2 s+ 3 s^2 +2(M_{U_1}^2+s)^2\log \left(\frac{M_{U_1}^2}{M_{U_1}^2+s}\right)\right) \right]\notag\\
&+\left. 4|\lambda_{q\mu}|^4\left( s(2+\frac{s}{M_{U_1}^2})+2(M_{U_1}^2+s)\log\left(\frac{M_{U_1}^2}{M_{U_1}^2+s}\right)\right)\right]~.
\end{align}

Analogously, we compute analytically all other $2 \to 2$ cross sections for the processes discussed in Section~\ref{sec:MuC}.
In the processes that include diagrams with a photon exchanged in the $t$-channel, such as $\mu^+ \mu^- \to \mu^+ \mu^-$ and $\mu q \to \mu q$ (see diagrams in Fig. \ref{fig:FDiagrams}), one cannot integrate over the whole phase space due to the pole in the propagator. Instead, we calculate numerically a fiducial cross-section by integrating over the physical region with a rapidity cut $|y_i|\lesssim 2$ on the final state fermions, using Eq.~\eqref{eq:triplediffXsec}.

\section{Statistical procedure}
\label{app:statistics}

To derive the expected exclusion or discovery reach we construct our test statistic as $-2 \log L = - 2  \sum_{i \in \text{bins}} \log L_i$, where
\be\begin{split}
  \text{if } N_i^{\rm obs} \geq 100~:& \qquad   -2 \log L_i = \frac{(N_i - N_i^{\rm obs})^2}{N_i + \epsilon^2 N_i^2}~, \\
  \text{if } N_i^{\rm obs} < 100~:& \qquad   -2 \log L_i = -2 \log \frac{N_i^{N_i^{\rm obs}} e^{- N_i}}{N_i^{\rm obs} !}~,
\end{split}\ee
where $N_i^{\rm obs}$ ($N_i$) is the observed (expected) number of events in each bin and $\epsilon$ is the relative systematic uncertainty. We assume that our test statistic follows a $\chi^2$ distribution.
In case of exclusion reach the observed number of events is derived assuming the SM, while the expected one is expressed in the new physics model. For the discovery reach, the expected number of events is given assuming the SM, while the observed one is derived assuming NP.

For the systematic uncertainty we assume an uncorrelated value of $\epsilon = 2\%$. This is increased to $10\%$ in the case of $\mu q \to \mu q$ process, due to our estimation of QCD uncertainties in quark PDFs inside the muon (see App.~\ref{app:MuonPDFs}).
The $2\%$ experimental systemic uncertainty is rather conservative according to some literature. For instance a value of 1\% is taken in \cite{AlAli:2021let}, while Refs.~\cite{Huang:2021nkl,Han:2020uak} assume a systematic uncertainty of only 0.1\%.

\section{Detector performance}
\label{app:resolution}
\label{app:Detector}

A detailed study of the FCC-hh detector system has been collected in the design report \cite{FCC:2018vvp}, from which we take the expected performances relevant for the processes considered in this paper.
Specifically, for the energy resolution of the hadronic calorimeter we take the baseline performance of the reference detector (c.f. Table 7.3):
\be
    \frac{\sigma_E}{E} = \frac{50\%}{\sqrt{E [\GeV]}} \oplus 3\%~,
    \label{eq:hCalRes}
\ee
where $\oplus$ means the two terms are added in quadrature.
For the muon $p_T$ resolution we take the combined resolution from the muon system and the tracker, assuming $25\mu m$ position resolution (c.f. Fig. 7.21(a) of Ref.~\cite{FCC:2018vvp}). This is $\approx 2\%$ at $1\,\TeV$ and $\approx 5\%$ at $10\,\TeV$.

For all MuCs considered, we assume that the hadronic calorimeter and muon system performances are the same as for the FCC-hh. However, when relevant, we limit the maximum rapidity to $|y|_{\rm max} = 2$, due to the rapid degradation of tracking efficiency closer to the beamline~\cite{Bartosik:2020xwr}. Our tracking and hadronic calorimeter resolutions are conservative with respect to the ones required for the CLIC detector~\cite{CLICdetector} in the corresponding energy range.

The LHC performance specifications are detailed in the CMS paper \cite{Sirunyan:2021khd}. The jet and muon triggering, identification, and reconstruction efficiencies are assumed to be 100\% in our analyses for FCC-hh and MuC.

\section{(HL-)LHC and FCC-hh analyses}
\label{app:FCChh}

\subsubsection*{$p p \to \mu^+ \mu^-$}

The analysis of the di-muon signatures at the hadron colliders is based on the recent CMS search \cite{CMS:2021ctt}.
For each benchmark model, we calculate the leading-order Drell-Yan cross section analytically on parton level which is then numerically convoluted with the {\tt NNPDF30\_nnlo\_as\_0118} PDF set \cite{NNPDF:2014otw} using Mathematica ($\mu_F = m_{\mu\mu}$).
The CMS collaboration reported the SM expected number of DY and other background events in the binned $m_{\mu\mu}$ distribution.
Their DY prediction is calculated at NNLO QCD and NLO EW precision. 

The present observed 95\% CL limits on all benchmark models come from assuming the number of expected DY events in each bin reported by the CMS collaboration is rescaled by the ratio of leading-order BSM and SM cross sections~\cite{Greljo:2017vvb}.
The systematic uncertainty on the number of DY events is as reported by the CMS collaboration.
The statistical analysis takes into account the number of SM+BSM DY events, as well as events coming from non-DY backgrounds.
We note, that neglecting the subleading contribution of non-DY backgrounds has a negligible effect on the derived bounds. 

In order to obtain the projections for the HL-LHC and the FCC-hh, we translated the $m_{\mu\mu}$ distribution into $\tau=m^2_{\mu\mu}/s_0$ distribution and rescaled the CMS prediction by the ratio 
\begin{equation}
    \sigma_i(s) = \sigma^{\rm SM\,CMS }_i(s_0) \frac{\sigma^{\rm SM\,LO}_i(s)}{\sigma^{\rm SM\,LO}_i(s_0)}\frac{L}{L_0}\,,
\end{equation}
where the index $i$ labels the bin, $s_{(0)}$ is the future (present) center-of-mass energy and $L_{(0)}$ is the future (present) luminosity.
The validity of such scaling for DY cross section was checked at LO using {\tt MadGraph5}~\cite{Alwall:2014hca}.
The non-DY backgrounds, however, do not scale in the same way.
While $t\bar t$ and $VV$ backgrounds remain subleading at the FCC-hh (directly checked by {\tt MadGraph5}), the $tW$ background becomes relevant (less than an order of magnitude times the SM DY). 
However, the presence of jets and missing energy in the final state allows for designing cuts that can easily suppress this background. We therefore neglect any non-DY backgrounds in the derivation of the future projections.

In all non-resonant searches, as well as resonant search at the LHC and the HL-LHC, the CMS binning of the $\tau$ distribution was used.
On the other hand, similarly to the MuC searches, the resonant search at FCC-hh uses bins constructed following the hadronic calorimeter resolution described in Eq.~\eqref{eq:hCalRes}.
For HL-LHC and FCC-hh we assign 2\% systematic uncertainty.

\subsubsection*{$p p \to 4 \mu$}

In the case of multilepton signature for $L_\mu - L_\tau$ model, we derive the exclusion limits and discovery reach by calculating the significance of the signal as 
\begin{equation}
    Z(M_{Z'},g_{Z'}) =\frac{s(M_{Z'},g_{Z'})}{\sqrt{b}} = \sqrt{L}~\frac{g^2_{Z'}\sigma_{Z'}(M_{Z'},1)}{\sigma_{\rm SM}}\,.
\end{equation}
where both cross sections were calculated using {\tt MadGraph5} applying the following set of standard cuts: $p_T^\mu > 20$ GeV (leading muon), $p_T^\mu > 10$ GeV (subleading muon), $p_T^\mu > 5$ GeV (other muons), $\left|\eta_\mu\right| < 2.7$ and $\Delta R_{\mu\mu}> 0.05$. Additionally, in order to suppress the SM background, we applied the cut $m_{\mu^+ \mu^-} > \min(800\ \mathrm{GeV}, 0.8 M_{Z'})$ on any oppositely-charged muon pair.

\bibliographystyle{JHEP}
\bibliography{Biblio}

\end{document}